\documentclass[12pt]{article}
\usepackage{epsf}
\usepackage{amsmath}
\usepackage{graphicx}
\usepackage{color}
\usepackage{psfrag}
\usepackage{amssymb}

\usepackage{ifpdf}

\newcommand{\bmat}{\left(\begin{array}}
\newcommand{\emat}{\end{array}\right)}

\def\yzero{\smash{\hbox{$y\kern-4pt\raise1pt\hbox{${}^\circ$}$}}}

\def\beq{\begin{equation}}
\def\eeq{\end{equation}}
\def\beqa{\begin{eqnarray}}
\def\eeqa{\end{eqnarray}}

\def\-{\hphantom{-}}
\def\ov{\overline}
\def\s2{\frac{1}{\sqrt2}}

\def\beq{\begin{equation}}
\def\eeq{\end{equation}}
\def\beqa{\begin{eqnarray}}
\def\eeqa{\end{eqnarray}}
\def\tr{{\rm tr \,}}
\def\Tr{{\rm Tr \,}}
\def\Pf{{\rm Pf \,}}
\def\nn{\nonumber}
\def\diag{{\rm diag \,}}
\def\IF{\relax{\rm I\kern-.18em F}}
\def\II{\relax{\rm I\kern-.18em I}}

\def\cn{{\cal N}}

\def\Dsl{\,\raise.15ex\hbox{/}\mkern-13.5mu D} 

\def\IS{\bf S}

\def\IZ{\bf Z}
\def\IT{\bf T}

\def\CN{{\cal N}}

\catcode`\@=11   
\newdimen\@rotdimen
\newbox\@rotbox  

\def\@vspec#1{\special{ps:#1}}
\def\@rotstart#1{\@vspec{gsave currentpoint currentpoint translate
   #1 neg exch neg exch translate}}
\def\@rotfinish{\@vspec{currentpoint grestore moveto}}
\def\@rotr#1{\@rotdimen=\ht#1\advance\@rotdimen by\dp#1%
   \hbox to\@rotdimen{\hskip\ht#1\vbox to\wd#1{\@rotstart{90 rotate}%
   \box#1\vss}\hss}\@rotfinish}
\def\@rotl#1{\@rotdimen=\ht#1\advance\@rotdimen by\dp#1%
   \hbox to\@rotdimen{\vbox to\wd#1{\vskip\wd#1\@rotstart{270 rotate}%
   \box#1\vss}\hss}\@rotfinish}%
\def\@rotu#1{\@rotdimen=\ht#1\advance\@rotdimen by\dp#1%
   \hbox to\wd#1{\hskip\wd#1\vbox to\@rotdimen{\vskip\@rotdimen
   \@rotstart{-1 dup scale}\box#1\vss}\hss}\@rotfinish}%
\def\@rotf#1{\hbox to\wd#1{\hskip\wd#1\@rotstart{-1 1 scale}%
   \box#1\hss}\@rotfinish}%
\def\rotate{\@ifnextchar[{\@rotate}{\@rotate[l]}}
\def\@rotate[#1]#2{\setbox\@rotbox=\hbox{#2}\@nameuse{@rot#1}\@rotbox}

\catcode`\@=12

\topmargin
-1.5cm
\textwidth
15.5cm
\textheight
23.5cm
\oddsidemargin
0.7cm
\evensidemargin
1.2cm

\begin{document}

\makeatletter
\@addtoreset{equation}{section}
\makeatother
\renewcommand{\theequation}{\thesection.\arabic{equation}}
\pagestyle{empty}
\rightline{ IFT-UAM/CSIC-10-02}
\vspace{0.1cm}
\begin{center}
\LARGE{\bf D-instanton and polyinstanton effects\\ from type I' D0-brane loops
\\[12mm]}
\large{Christoffer Petersson$^\dagger$, Pablo Soler$^\ddagger$, Angel M. Uranga$^{\ddagger \, *}$\\[3mm]}
\footnotesize{$^\dagger$ Department of Fundamental Physics\\ Chalmers University of Technology, 412 96 G\"oteborg, Sweden}\\
\footnotesize{$^\ddagger$Instituto de F\'{\i}sica Te\'orica UAM/CSIC,\\[-0.3em]Universidad Aut\'onoma de Madrid C-XVI, Cantoblanco, 28049 Madrid, Spain} \\[2mm] 
\footnotesize{$^*$ PH-TH Division, CERN CH-1211 Geneva 23, Switzerland}\\

\vspace{1.3cm}
\small{\bf Abstract} 
\end{center}
\begin{center}
\begin{minipage}[h]{16.0cm}
We study non-perturbative D$(-1)$-instanton corrections to quartic gauge and curvature couplings in 8d type IIB orientifolds, in terms of a one-loop computation of BPS D0-branes in T-dual type I' models. The complete perturbative and non-perturbative results are determined by the BPS multiplicities of perturbative open strings and D0-brane bound states in the 9d type I' theory. Its modular properties admit a geometric interpretation by lifting to Horava-Witten theory. We use the type I' viewpoint to motivate an interpretation of 8d and 4d polyinstanton effects, consistent with heterotic - type II orientifold duality.

\end{minipage}
\end{center}
\newpage
\setcounter{page}{1}
\pagestyle{plain}
\renewcommand{\thefootnote}{\arabic{footnote}}
\setcounter{footnote}{0}

\tableofcontents


\section{Introduction}

The systematic computation of non-perturbative D-brane instanton effects is an important question in string compactifications \cite{Becker:1995kb,Witten:1996bn,Harvey:1999as,Witten:1999eg,Billo:2002hm,Kiritsis:1999ss}, with potentially interesting phenomenological implications \cite{Blumenhagen:2006xt,Ibanez:2006da}. Among several recent developments (see \cite{Blumenhagen:2009qh} for a review with an extensive list of references), an interesting direction is the extension of gauge theory localization techniques \cite{Moore:1998et} to general D-brane instanton effects, in particular D$(-1)$-brane instanton (and multi-instanton) corrections to $F^4$ couplings on D7-branes in type IIB orientifolds \cite{Billo:2009di,Fucito:2009rs}. 

The understanding of D-brane instanton effects in orientifold models is important, and it would be desirable to have alternative viewpoints on these results. In this paper we recover and generalize these results using a different technique. Following \cite{Ooguri:1996me} one can resum multiinstanton effects by computing one-loop diagrams in a T-dual theory, in which the instantons turn into D-brane particles running in a loop winding along the T-dual circle. This picture has appeared in several contexts,  e.g. \cite{Gaiotto:2008cd,Collinucci:2009nv}, to provide insight into the continuity of non-perturbative effects across lines of BPS stability of the microscopic instantons \cite{GarciaEtxebarria:2007zv,GarciaEtxebarria:2008pi}. In this paper, we use it as a computational tool, by which we obtain the $F^4$ and $R^4$ instanton corrections to an 8d type IIB orientifold from a one-loop computation with the spectrum of BPS particles in a T-dual type I' configuration. The computation is extremely simple, and the result is determined in terms of well-studied degeneracies and quantum numbers of 9d type I' BPS states. Our analysis is related to the ideas in \cite{Gutperle:1999dx}.

Besides recovering and extending known results, our work has an interesting spinoff. The type I' computation allows for a direct comparison with the dual heterotic worldsheet instanton computation, since both can be regarded as one-loop diagrams of BPS particles. This comparison allows us to revisit the proposal in \cite{Blumenhagen:2008ji} of the existence and nature of certain polyinstanton effects, which also exist in these 8d models. 
Our analysis suggests that these effects do not conflict with heterotic-type I duality. Microscopically they can be regarded as reducible diagrams not correcting the microscopic 1PI action. Contributions to the low-energy 1PI effective action, or to the Wilsonian action, can thus be regarded as generated by suitable spacetime tree level diagrams of single instanton vertices.

The paper is organized as follows: In Section \ref{computation} we compute D$(-1)$-brane instanton corrections in type IIB orientifolds with D7-branes/O7-planes as one-loop diagrams of BPS particles in type I'. In section \ref{background} we remind the reader of this setup and in section \ref{generalities} we show the general structure of the one-loop amplitudes. In sections \ref{sosixteen} and \ref{soeight} we apply them to the computation of instanton corrections in 8d models with $SO(16)^2$ and $SO(8)^4$ gauge symmetry. In 
section \ref{perturbative} we show that the 8d type IIB perturbative $F^4$ corrections can be obtained from similar computations, using perturbative BPS particles in type I'. In section \ref{gravitational} we compute gravitational $R^4$ and mixed $R^2F^2$ corrections with these techniques. 

In Section \ref{moregeneral} we consider more general configurations. In section \ref{wilsonpositions} we consider general Wilson lines and positions for the type I' D8-branes, corresponding to general positions of the type IIB D7-branes in the transverse complex plane. The general result is efficiently packaged in an 8d prepotential.
In section \ref{generalgravitational} we extend these ideas to the computation of gauge and gravitational corrections.

In Section \ref{mtheory} we recover these results from the perspective of Horava-Witten M-theory, via a computation of one-loop diagrams for KK momentum modes of bulk gravitons and boundary $E_8$ gauge bosons, which make the modular properties manifest.

In Section \ref{polyinstanton} we discuss polyinstanton effects. In section \ref{8dpoly} we introduce 8d polyinstantons, in analogy with the 4d polyinstantons in \cite{Blumenhagen:2008ji}. In section \ref{polyduality} we show that type IIB polyinstanton effects are not included in the standard worldsheet instanton corrections of the heterotic dual. 
In section \ref{poly1pi} we argue that there is no conflict with duality since polyinstanton processes do not contribute to corrections to the microscopic 1PI action, but rather represent reducible Feynman diagrams.

In Section \ref{compactification} we describe the basic features of D$(-1)$-brane instanton corrections in orientifolds of type IIB on K3$\times \IT^2$, and their relation to one-loop diagrams in the T-dual type I' compactification. Since these contributions can be regarded as a simple dimensional reduction, all conclusions about 8d polyinstantons remain valid for polyinstanton corrections to gauge kinetic functions in the resulting 4d $\CN=2$ models (and further reductions to $\CN=1$ by freely acting quotients).  

We present our conclusions in Section \ref{conclusions}, and leave some details of group theoretical traces and anomaly polynomials for appendices \ref{traces} and \ref{anomaly}.

\section{The type I' computation}
\label{computation}

\subsection{The background}
\label{background}

The background we consider is 9d type I' theory \cite{Polchinski:1995df}, namely type IIA on $\IS^1$ modded out by the orientifold action $\Omega R (-1)^{F_L}$, where $R$ flips the circle coordinate and $F_L$ is the left-moving fermion number. There are two O8-planes and 32 D8-branes to cancel the RR tadpole. We work in the covering space picture. 

Throughout the paper we focus on the configuration with 16 D8-branes on top of each O8-plane, hence the 9d gauge symmetry is $SO(16)^2$. There is local RR tadpole cancellation and constant dilaton profile along the circle. The spectrum of 9d BPS one-particle states will be important for us: there are perturbative open string BPS states stretching among the D8-branes, hence carrying integer winding charge for states in the $({\bf 120,1})+({\bf 1,120})$, and with half-integer winding for states in the $({\bf 16},{\bf 16})$. In addition, there are non-perturbative BPS particle states arising from D0-brane bound states. For D0-brane bound states in the bulk, there is a single BPS state with $2n$ units of D0-brane charge (as counted in the covering space), and neutral under the gauge symmetry. For D0-branes on top of each O8-plane, there is a single BPS state with charge $2n$, transforming in the ${\bf 120}$ of the corresponding $SO(16)$, and a single BPS state with charge $2n+1$ 
 in the ${\bf 128}$. These BPS multiplicities can be simply understood by regarding these states as KK momentum modes of the gravity multiplet and the $E_8$ vector multiplets in the lift to Horava-Witten theory. The different transformation properties of the boundary D0-brane states is due to the momentum shift from the Wilson lines in the eleventh direction, breaking $E_8\to SO(16)$, see e.g. \cite{Kachru:1996nd}. More general type I' configurations and their BPS spectrum have been considered in \cite{Bergman:1997gf,Bergman:1997py}, whose results will be useful in some generalizations.

In this paper we consider the compactification of this 9d model to 8d on a further $\IS^1$, along which we may have general $SO(16)^2$ Wilson lines. This model is related by T-dualtity to the type IIB orientifold in \cite{Billo:2009di}, where towers of D$(-1)$-brane instantons generate contributions to couplings $F^4$ and $R^4$. In a dual heterotic model, these are generated as one-loop threshold corrections. These corrections have been studied from different viewpoints, see e.g. \cite{Bachas:1997mc,Kiritsis:1997hf,Kiritsis:1999ss,Lerche:1998nx,Lerche:1998gz,Lerche:1998pf,Foerger:1998kw,Gava:1999zk}. Our purpose is to determine these couplings in the type I' picture where they can be easily described in terms of a one-loop diagram of 9d BPS particles, possibly winding around the $\IS^1$ compactification circle to 8d. Perturbative BPS open strings and non-perturbative D0-brane BPS bound states will produce contributions T-dual to perturbative and D$(-1)$-instanton contribution
 s in
  the type IIB orientifold.

\subsection{Generalities of one-loop diagrams}
\label{generalities}

The computations in this subsection follow \cite{Gutperle:1999dx}.
We are interested in computing the one-loop diagram of a 9d BPS particle on $M_8\times \IS^1_9$, with four external insertions of gauge fields strengths (or curvatures). For concreteness we focus on D0-branes, although the result is more general, see section \ref{perturbative}. The amplitude has the following structure,
\begin{equation}
\label{amp}
\mathcal{A}_{F^4}= \int_{0}^{\infty} \frac{dt}{t}\sum_{\ell_9} \int d^{8}\mathbf{p} ~ e^{-t \left( \mathbf{p}^2+\frac{(\ell_9-\tilde{c})^2}{R^{2}_{9}} +\mu^2 \right)} \Tr \left[ \prod_{r=1}^{4} \int_{0}^{t} d\tau_{r} V_{F}^{(r)} (\tau_r )  \right]
\end{equation}
where the sum is over KK-momenta $\ell_9$ and $\mu$ is the 9d mass of the BPS particle. The Wilson line $\tilde{c}=c+A$ along $\IS^1_9$, has a piece $c$ from a gauge boson in the gravity multiplet (i.e. the RR 1-form for BPS D0-branes) and a piece $A$ encoding possible Wilson lines of vector multiplet gauge bosons (i.e. the D8-brane $SO(16)$ Wilson lines for D0-branes). 

For BPS protected amplitudes, the vertex operator part can simply be replaced by a factor of $t^4$ and insertions encoding the coupling to the external gauge bosons or curvatures. Focusing on the former case, we obtain an insertion of $\tr_{\mathbf{R}} F^4$ for a BPS particle in a representation ${\mathbf{R}}$ of the gauge group. Corrections involving curvatures are discussed in section \ref{gravitational}.

The amplitude is a particular case $d=9$, $k=4$ of the expression for the amplitude in $d$ dimensions with $k$ insertions
\begin{equation}
\mathcal{A}=\frac{1}{k!\,\pi^{(d-1)/2}} \sum_{\ell_9} \int d^{d-1}\mathbf{p} \int_{0}^{\infty} \frac{dt}{t} ~t^k ~e^{-t \left( \mathbf{p}^2 + \frac{(\ell_9-\tilde{c})^2}{R_9^2} +\mu^2 \right)}~.
\label{generalamp}
\end{equation}
Integrating over continuous momenta 
and performing a Poisson resummation over the discrete one, we get
\begin{equation}
\label{gral}
\mathcal{A}=\frac{\sqrt{\pi}\,R_9}{k!} \sum_{w_9}  \int \frac{dt}{t} ~t^{k-d/2} ~e^{-\frac{\pi^2 R_9^2 w_9^2}{t}-t \mu^2 }e^{-2\pi i w_9 \tilde{c}}~.
\end{equation}
We will focus in this section on the case $w_9\neq0$ (states with $w_9=0$ lead to a tree-level contribution which will be considered in section \ref{perturbative}). Using the integral representation of Bessel functions
\begin{equation}
\label{ }
\int_{0}^{\infty} \frac{dt}{t}t^{-s} e^{-\frac{A}{t}-tB}=2\left| \frac{B}{A} \right|^{s/2} K_s \left( 2\sqrt{|AB|} \right)
\end{equation}
we can rewrite the amplitude as
\begin{equation}
\label{1loop}
\mathcal{A}_{(d/2-k)}=\frac{2\sqrt{\pi}\,R_9}{k!} \sum_{w_9\neq0}  \left| \frac{\mu}{\pi R_9 w_9} \right|^{d/2-k} K_{d/2-k} \left( |2\pi R w_9 \mu| \right) e^{-2\pi i w_9 \tilde{c}}~
\end{equation}
where we have indicated on $\mathcal{A}$ that the amplitude depends on the number $(d/2-k)$. It is interesting to point out that using the differentiation formula  for Bessels functions (10.2.22) in \cite{aands}, one can write
\beqa
\mathcal{A}_{d,k} (\mu) & = &  \left(\, -\frac 1\mu\,\frac{d}{d\mu} \,\right)^k\, \mathcal{A}_{d,k=0} (\mu) 
\eeqa
namely the $k$-leg amplitude can be obtained as the $k^{th}$ derivative of the vacuum amplitude. In section \ref{moregeneral} the amplitudes with external legs will indeed be generated from a Schwinger-like vacuum amplitude in a general background, by differentiation with respect to background vector multiplets, on which the BPS masses depend.

For our case of interest $d/2-k=1/2$, the saddle point approximation is exact, $K_{1/2}(x)=\sqrt{\frac{\pi}{2x}}e^{-x}$, and we get
\begin{eqnarray}
\label{1loop1/2}
\mathcal{A}_{(1/2)}
&=&\frac{1}{4!}\sum_{w_9\neq0} \frac{1}{|w_9|}~e^{-2\pi R_{9} |w_9 \mu|-2\pi i w_9 \tilde{c}}~.
\end{eqnarray}
For a bound state of $n$ D0-branes, $\mu=n/g_s$ and $c=nc_0$, with $c_0$ the RR 1-form Wilson line. Defining  $q=e^{2\pi i \tau}$ with $\tau= i \frac{R_{9}}{g_s}+c_0$, 
its contribution is
\beqa
\Delta^{D0}=\frac{1}{4!}\sum_{w_9>0}\,\frac{1}{w_{9}}\, q^{n\,w_9}\, e^{2\pi i \, w_9A}\, \tr_{\mathbf{R}} F^4 ~ + \mathrm{c.c.} 
\label{master}
\eeqa
This expression already suggests an instanton expansion. Notice that the prefactor is simply constant, as expected for the high supersymmetry of the system.

This basic expression allows for the computation of instanton corrections to 8d theories obtained upon further circle compactification to 8d, possibly with Wilson lines. In the following we work out several examples of gauge and gravitational corrections. In the latter case, suitable curvature traces replace $\tr F^4$ in the expression.

\subsection{The $SO(16)$ model}
\label{sosixteen}

Consider the simplest situation where there are no Wilson lines on the circle $\IS^1_9$, so that the 8d gauge group is $SO(16)^2$. The non-perturbative gauge threshold corrections are easily computed from the D0-brane one-loop diagrams, using the BPS bound state spectrum information: For each boundary there exists a D0-brane BPS bound state of mass $|2n|$, in the representation ${\bf{120}}$ of the corresponding $SO(16)$, and one state of mass $|2n-1|$ in the ${\bf{128}}$, for each non-zero integer $n$. Focusing on a single $SO(16)$, the contribution is
\begin{eqnarray}
\Delta^{D0}_{SO(16)}&= &\frac{2}{4!} \sum_{w_9,n>\,0}\,  \frac{1}{w_9}\, q^{w_9(2n)}\, \tr_{\mathbf{ 120}} F^{4}\, +\frac{2}{4!} \, \sum_{w_9,n}\,  \frac{1}{w_9}\, q^{w_9(2n-1)}\, \tr_{\mathbf{ 128}} F^{4} \, + \mathrm{c.c.} \, \, \nn \\
& = & 
\frac{1}{3} \, \tr F^{4} \,\sum_{k}\sum_{\ell \large| k} \frac{1}{\ell}\, \left[ 2 q^{2k} - q^k \right]\, +\,  \frac{1}{8}\,( \tr F^{2})^2 \, \sum_{k}\sum_{\ell \large| k} \frac{1}{\ell} \,\left[ - q^{2k} + 2 q^k \right] \, +\mathrm{c.c.} \nn\\
&&
\end{eqnarray}
where in the last equality we have used (\ref{SO16}) to rewrite the expression in terms of traces in the vector representation. This result agrees with the result in \cite{Gutperle:1999dx} for heterotic strings on $\IT^2$. 

\subsection{The $SO(8)^2$ model}
\label{soeight}

Let us now consider turning on $\IZ_2$ valued Wilson lines which break each 9d $SO(16)$ factor to $SO(8)^2$. This model was discussed in \cite{Billo:2009di}. The 9d BPS states  in $SO(16)$ representations pick up different phases according to their behaviour under the decomposition
\begin{eqnarray}
&&  \mathbf{ 120}  \to  (\mathbf{ 28},\mathbf{ 1})_{+}+(\mathbf{ 1},\mathbf{ 28})_{+}+(\mathbf{ 8_v} ,\mathbf{ 8_v})_{-} \nn\\
&& \mathbf{ 128} \to (\mathbf{ 8_s},\mathbf{ 8_s})_{+}+(\mathbf{ 8_c},\mathbf{ 8_c})_{-} 
\label{SO(8)reps}
\end{eqnarray}
where the subindex $\pm$ corresponds to having $e^{2\pi i\,A}=\pm 1$.

We focus on $F^4$ correction associated to only one of the $SO(8)$ factors, in which case only states charged under this factor contribute. The result is (we will drop the $+\,\mathrm{c.c.}$ term in the following)
\beqa
\Delta_{SO(8)}^{D0} &=& \frac{2}{4!}\,\Bigl[\sum_{n,w_9}\, \frac{1}{w_9}\, q^{2n\, w_9}\, \tr_{\mathbf{28}}\,F^4\, +\, 8\, \sum_{n,w_9}\, \frac{1}{w_9}\, q^{2n\, w_9}\, (-1)^{w_9}\, \tr_{\mathbf{8_v}}\,F^4\, \nonumber \\
&&~~ +\,8\, \sum_{n,w_9}\, \frac{1}{w_9}\, q^{(2n-1)\, w_9}\, \tr_{\mathbf{8_s}}\,F^4\, +\, 
8\, \sum_{n,w_9}\, \frac{1}{w_9}\, q^{(2n-1)\, w_9}\,(-1)^{w_9}\,  \tr_{\mathbf{8_c}}\,F^4\Bigr]~. \nn \\
\eeqa
Using the trace identities (\ref{traceso8}), we obtain
\beqa
\Delta_{SO(8)}^{D0} &=& \frac{1}{3}\,  \tr F^{4}\, \sum_{w_9,n} \, \left[ \, \frac{1}{2w_9}\, q^{(2n)\,2w_9}\, -\,\frac{1}{2w_9-1}\, q^{(2n)\,(2w_9-1)}\, -\,\frac{1}{2w_9}\, q^{(2n-1)\,2w_9}  \,\right]\,  \nn \\
&&+  \frac{1}{8}\, (\tr F^2)^2\, \sum_{n,w_9}\, \left[\, \frac{1}{w_9}\, q^{2n\, w_9}\, +\, 2\times \frac{1}{2w_9}\, q^{(2n-1)\, 2w_9}\, \right]\, \nn \\
&&-\, 8\, \Pf F\, \sum_{n,w_9}\, \frac{1}{2w_9-1}\, q^{(2n-1)(2w_9-1)}\,  \nn\\
&=& \frac{1}{2} \, \tr F^4\,  \sum_{k}\sum_{\ell \large| k} \, \frac{1}{\ell}\, \left[ \, q^{4k}\, -\, q^{2k}\, \right]\, -\, \frac{1}{8}\, (\tr F^2)^2\, \sum_{k}\sum_{\ell \large| k} \, \frac{1}{\ell} \, \left[ \, q^{4k}\, -\, 2q^{2k}\, \right]\,  \nn\\
&&- 8\, \Pf F\, \sum_{k}\sum_{\ell \large| 2k-1}\, \frac{1}{\ell}\, q^{2k-1}~.
\eeqa
This agrees with the correction in \cite{Billo:2009di}, up to an overall minus sign.

Note that in principle the D0-branes could seem to generate mixed corrections $\tr F_1^2\, \tr F_2^2$ for two $SO(8)$ factors from the same boundary. However these vanish due to a cancellation between D0-branes in different representations. Explicitly,
\beqa
\label{zeromixed}
&& \sum_{n,w_9}\, \frac{1}{w_9}\, q^{2n\, w_9}\, (-1)^{w_9}\, \tr_{\mathbf{8_v}}\, F_1^{\, 2}\, \tr_{\mathbf{8_v}}\, F_2^{\, 2}\, +\, 
\sum_{n,w_9}\, \frac{1}{w_9}\, q^{(2n-1)\, w_9}\, \tr_{\mathbf{8_s}}\, F_1^{\, 2}\, \tr_{\mathbf{8_s}}\, F_2^{\, 2}\, \nn\\
& &+ \sum_{n,w_9}\, \frac{1}{w_9}\, q^{(2n-1)\, w_9}\, (-1)^{w_9}\, \tr_{\mathbf{8_c}}\, F_1^{\, 2}\, \tr_{\mathbf{8_c}}\, F_2^{\, 2}\, \\
&=& \tr F_1^2\, \tr F_2^2\, \left[\, \sum_{n,w_9}\, \frac{1}{2w_9}\, q^{2n\,2w_9}\, -\,\sum_{n,w_9}\, \frac{1}{2w_9-1} \, q^{2n\,(2w_9-1)}\, +\, 2\times \frac{1}{2w_9}\, q^{(2n-1)\, 2w_9}
\,\right]\, =0\nn
\eeqa
where we have used (\ref{traceso8}) and the cancellation follows after some simple manipulations.

\subsection{Perturbative contributions}
\label{perturbative}

Perturbative corrections to the gauge couplings we are computing arise in two different ways. First of all, bound states of D0 branes with zero winding in the circle $\IS_9$ yield a tree-level contribution which can be easily obtained from \eqref{gral} by setting $w_9=0$: 
\begin{eqnarray}
\label{tree-level}
\Delta^{\mathrm{Pert}}&=&\frac{\sqrt{\pi}\,R}{4!} \sum_{n\in \mathbb{Z}}  \int \frac{dt}{t} ~t^{-\frac{1}{2}} \left[e^{-t \frac{(2n)^2}{g_s^2}}\,\tr_{\mathbf{120}} F^4 +
e^{-t \frac{(2n-1)^2}{g_s^2}}\,\tr_{\mathbf{128}} F^4 \right] \\
&=& \frac{4 \tau_2}{4!\, \pi} \sum_{w>0}\left[ \frac{1}{w^2} \tr_{\mathbf{120}}F^4 + \frac{(-1)^{w}}{w^2} \tr_{\mathbf{128}}F^4\right]= \frac{\pi\, \tau_2}{3\cdot4!} \left[ 2\,\tr_{\mathbf{120}}F^4 - \tr_{\mathbf{128}}F^4\right]~. \nonumber
\end{eqnarray}
Using the trace identities of appendix \ref{traces}, we can calculate these contributions for the $SO(16)^2$ and the $SO(8)^4$ models:
\begin{eqnarray}
&\Delta_{SO(16)}^{\mathrm{Pert}}&=\frac{\tau_2\, \pi}{6}\,\tr F_{SO(16)}^4~~~~~~~~~~~\nonumber\\
&\Delta_{SO(8)}^{\mathrm{Pert}}&=\frac{\tau_2\, \pi}{6}\,\tr F_{SO(8)}^4~.
\end{eqnarray}
Again, these terms agree with the heterotic results in the literature.

\medskip

The spectrum of BPS particles in the 9d type I' theory also contains perturbative states, corresponding to open strings winding in the interval $\IS^1_{10}/\IZ_2$. Since the BPS condition forbids any oscillation excitation, they are simply the groundstates of open strings stretching between the D8-branes. There are states starting and ending on the same $SO(16)$ stack, therefore transforming in the corresponding ${\mathbf{120}}$ and labeled by an integer winding $w_{10}$, and states stretching between the two $SO(16)$ stacks, thus in the $(\mathbf{16},\mathbf{16})$ and labeled by a half-integer winding $w_{10}-1/2$. In the $\IS^1_9$ compactification to 8d, one-loop contributions from these states can be analyzed using formulas similar to the above, by simply taking into account their different 9d masses. The results correspond to perturbative one-loop $F^4$ terms. 

The starting point is the analog of the amplitude \eqref{amp}, for perturbative states:
\begin{eqnarray}
\mathcal{A}^{Pert}&=&\frac{1}{\pi^{8} 4!} \sum_{{}^{\ell_9\in\mathbf {Z}}_{w_{10}\in\frac{\mathbf{Z}}{2}}} \int d^{8}\mathbf{p} \int_{0}^{\infty} \frac{dt}{t} ~t^4 ~e^{-t \left( \mathbf{p}^2 + \frac{(\ell_9-\tilde{b})^2}{R_{9}^2} +w_{10}^2R_{10}^2 \right)} \nonumber \\
&=& \frac{\sqrt{\pi} R_9}{4!} \sum_{{}^{w_9\in\mathbf {Z}}_{w_{10}\in\frac{\mathbf{Z}}{2}}} \int_{0}^{\infty} \frac{dt}{t} ~t^{-1/2} ~e^{-\frac{\pi^2 R_9^2 w_9^2}{t}} ~e^{-t w_{10}^2 R_{10}^2} ~e^{-2\pi i w_9\tilde{b}}~,
\label{amppert}
\end{eqnarray}
where we have integrated over continuous momenta and performed a Poisson 
resummation over the discrete momentum $\ell_9$. Here, $\tilde{b}$ contains the 
NSNS 2-form and the information about the Wilson line along $\mathbf{S}^1_9$, 
in complete analogy with $\tilde{c}$ for the non-perturbative 
contributions: $\tilde{b}=b+A=w_{10} b_0 + A$.

Let us first consider the contributions from states with non-zero winding in 
the interval $\IS^1_{10}/\IZ_2$, but with $w_9=0$, the analog of \eqref{tree-level}: 
\begin{eqnarray}
\Delta^{\mathrm{F1}}_{w_9=0,w_{10}\neq0}&=& \frac{\sqrt{\pi} R_9}{4!} \sum_{w_{10}\in\mathbf{Z}} \int_{0}^{\infty} \frac{dt}{t} ~t^{-1/2} ~e^{-t w_{10}^2 R_{10}^2}~\tr_{\mathbf{120}} F^4 \nonumber \\
&&+ \frac{\sqrt{\pi} R_9}{4!} \sum_{w_{10}\in\mathbf{Z}} \int_{0}^{\infty} \frac{dt}{t} ~t^{-1/2} ~e^{-t (w_{10}-1/2)^2 R_{10}^2}~\tr_{\mathbf{(16,16)}} F^4
\nonumber \\
&=& \frac{2U_2}{\pi 4!}\sum_{\ell_{10}>0}\frac{1}{\ell_{10}^2}~\tr_{\mathbf{120}} F^4 
+ \frac{2U_2}{\pi 4!}\sum_{\ell_{10}>0}\frac{(-1)^{\ell_{10}}}{\ell_{10}^2}~\tr_{\mathbf{(16,16)}} F^4 \nonumber \\
&=& \frac{\pi U_2}{3\cdot4!}~\tr_{\mathbf{120}} F^4 -\frac{\pi U_2}{6\cdot4!}~\tr_{\mathbf{(16,16)}} F^4
\label{zerow9}
\end{eqnarray}
where we performed a Poisson resummation and omitted the (divergent) term $\ell_{10}=0$. 
Here, $U=b_0+iR_9 R_{10}$ is the volume modulus of the compactification torus (in the covering 
space) which upon T-duality becomes the complex structure modulus of type IIB.

States with both winding numbers $w_9$ and $w_{10}$ different from zero contribute to the amplitude in a way 
equivalent to stacks of D0 branes with non-zero winding number along $\IS^1_9$ (see eq.\eqref{master}). 
The result is
\begin{eqnarray}
\Delta^{\mathrm{F1}}_{w_9\neq0,w_{10}\neq0}&=&\frac{2}{4!}\sum_{w_9,w_{10}>0}\, \frac{1}{w_9}\, q'^{\,w_9\, w_{10}}\,e^{-2\pi i w_9 A}\, \tr_{\mathbf{120}} F^4\, + \,\mathrm{c.c} \nonumber\\
&&+ \frac{2}{4!} \sum_{w_9,w_{10}>0}\, \frac{1}{w_9}\, q'^{\, w_9\,(w_{10}-\frac 12)}\,e^{-2\pi i w_9 A}\, \tr_{\mathbf{(16,16)}} F^4 +\, \mathrm{c.c.}\,,
\label{windings}
\end{eqnarray}
where we have defined $q'=e^{2\pi i U}$.

Finally there are contributions from states in the $\mathbf{120}$ with zero winding in the interval, 
corresponding to the massless gauge bosons. These states can be considered from eq.\eqref{amppert} 
by taking $w_{10}=0$ and $w_9\neq0$, or equivalently, from eq.\eqref{amp} by taking $\mu=0$:
\begin{eqnarray}
\Delta^{\mathrm{F1}}_{w_9\neq0,w_{10}=0}&=& \frac{\sqrt{\pi} R_9}{4!} \sum_{w_{9}\in\mathbf{Z}} \int_{0}^{\infty} \frac{dt}{t} ~t^{-1/2} ~e^{-2\pi i w_9 A-\frac{\pi^2R_9^2w_9^2}{t}}~\tr_{\mathbf{120}} F^4 \nonumber \\
&=& \frac{2}{4!}\sum_{w_9>0}\frac{e^{-2\pi i A w_9}}{w_9}~\tr_{\mathbf{120}} F^4.
\label{div}
\end{eqnarray}

\medskip

Let us consider the $SO(16)^2$ theory by setting $A=0$ in the above equations. In this case the contribution in \eqref{div} is divergent and the reason is because it arises from massless states running in the loop. This divergence may however be regularized\footnote{For further discussions on these issues, see section \ref{modular}.} using the prescription described in \cite{Dixon:1990pc}.  For our case, this prescription boils down to simply adding a term \cite{Green:1997as}, $\log(\tau_2U_2/\Lambda^2)$, which is consistent with modular invariance and where $\Lambda^2$ is chosen such that  it cancels the logarithmic divergence in the sum. By using this procedure we obtain
\begin{equation}
\Delta^{\mathrm{F1}}_{w_9\neq0,w_{10}=0}\to-\frac{1}{4!}\,\log(\tau_2U_2)\,\tr_{\mathbf{120}}F^4.
\label{regularized}
\end{equation}
The total contribution from open strings in this model is the sum of the terms \eqref{zerow9}, 
\eqref{windings} and \eqref{regularized}. Taking into account the trace identities of appendix \ref{traces} 
we obtain
\begin{eqnarray}
\Delta^{\mathrm{F1}}_{SO(16)}&=& \frac{-3}{4!}(\tr F^2)^2 \log(\tau_2U_2|\eta(U)|^4) \nonumber \\
&&+ \frac{8}{4!} \tr F^4 \, \left[-\log\,(\tau_2 U_2) + 2\sum_{w_9,w_{10}>0} (q'^{w_9w_{10}}+2\,q'^{w_9(w_{10}-\frac 12)}\, + \mathrm{c.c.})\right] \nn \\
\end{eqnarray}

\medskip

For the $SO(8)^4$ case we simply have to take into account the effect of the Wilson lines on 
each state and the breaking of the representations 
\begin{eqnarray}
  \mathbf{ 120}  &\to&  (\mathbf{ 28},\mathbf{ 1})_{+}+(\mathbf{ 1},\mathbf{ 28})_{+}+(\mathbf{ 8_v} ,\mathbf{ 8_v})_{-} \\
 (\mathbf{16};\mathbf{16}) &\to& (\mathbf{ 8_v,1};\mathbf{ 8_v,1})_{+}\, +\,  (\mathbf{ 8_v,1};\mathbf{1, 8_v})_{-}\, +\,  (\mathbf{1, 8_v};\mathbf{ 8_v,1})_{-}\, +\, 
 (\mathbf{1, 8_v};\mathbf{1, 8_v})_{+} \nn
\end{eqnarray}
Note that in this case \eqref{div} is only divergent for states in the 
$(\mathbf{28,1})_+$ and not for those in the $(\mathbf{8_v,8_v})_-$ for which $e^{2\pi i A}= -1$. 
For the latter, the sum in \eqref{div} yields just a moduli independent contribution which we will not consider. 
It turns out that the sum of the contributions to $\tr F^4$ from the $(\mathbf{ 8_v} ,\mathbf{ 8_v})_{-}$ states with non-zero winding cancel the contributions from the $(\mathbf{ 8_v,1};\mathbf{ 8_v,1})_{+}$ and $(\mathbf{ 8_v,1};\mathbf{1, 8_v})_{-}$ with non-zero winding. Explicitly,
\begin{equation}
\sum_{w_9,w_{10}}\, \frac{1}{w_9}\, q'^{\,w_9\, w_{10}}\, (-1)^{w_9}\, +\, 
\sum_{w_9,w{10}}\, \frac{1}{w_9}\, q'^{\,(w_{10}-\frac 12)\, w_9}\, +
\sum_{w_9,w_{10}}\, \frac{1}{w_9}\, q'^{\,(w_{10}-\frac 12)\, w_9}\, (-1)^{w_9} =0
\label{zeropert}
\end{equation}
where the cancellation is (for no obvious physical reason) analogous to that in (\ref{zeromixed}), as can be easily checked by trading $q'\to q^2$, $w\to n$. 

Finally, the contributions to $\tr F^4$ from states with zero winding in $\IS^1_9$ (eq.\eqref{zerow9}) also vanishes and therefore we conclude that this term does not receive (moduli dependent) contributions from fundamental strings (in the $SO(8)^4$ model).
Open strings in the $\mathbf{28}$ do however have a non-vanishing contribution to the $(\tr F^2)^2$ term. By collecting all the pieces we obtain,
\begin{eqnarray}
\Delta^{\mathrm{F1}}_{SO(8)}&=& \frac{1}{4!} 
\left[ \frac{\pi U_2}{3} - \log \left( \tau_2 U_2 \right) + 2\sum_{w,m}\, \frac{1}{m}\, \left[ q'^{\, wm}+\mathrm{c.c.} \right]  \right] \tr_{\mathbf{28}} F^4 \nn \\
&=& - \frac{3}{4!}\, \log \left( \tau_2 U_2 |\eta(U)|^4 \right) (\tr F^2)^2 
\label{pertf4so8}
\end{eqnarray}
which agrees with the heterotic result in \cite{Billo:2009di}. This result is also recovered in the type IIB model in \cite{Billo:2009di}, even though it is there treated as a local model. The agreement follows because of the cancellation (\ref{zeropert}) for contributions from open strings charged under different gauge factors.\\

There are also additional perturbative contributions to mixed terms like $\tr F_i^2\tr F_j^2$, which can be easily computed. Skipping the details, we get that
\beqa
\Delta_{SO(8)}^{\mathrm{Mixed}}&=& \frac{1}{4!} \sum_{w,m}\, \frac{1}{m}\, q'^{wm}\, (-1)^m \, [\, \tr F_1^{\,2}\, \tr F_2^{\, 2} \, +\, \tr F_3^{\,2}\, \tr F_4^{\, 2}\, ]\, \nn \\
&&+ \frac{1}{4!} \sum_{w,m}\, \frac{1}{m}\, q'^{(w-\frac 12)\, m}\, \, [\, \tr F_1^{\,2}\, \tr F_3^{\, 2} \, +\, \tr F_2^{\,2}\, \tr F_4^{\, 2}\, ]\, \nn \\
&&+ \frac{1}{4!} \sum_{w,m}\, \frac{1}{m}\, q'^{(w-\frac 12)\, m}\, (-1)^m \, [\, \tr F_1^{\,2}\, \tr F_4^{\, 2} \, +\, \tr F_2^{\,2}\, \tr F_3^{\, 2}\, ]
\eeqa 
These corrections have, to our knowledge, not been computed before for the heterotic or the type IIB dual.

\medskip

It is interesting that both perturbative and non-perturbative contributions can be discussed on an equal footing in the language of one-loop diagrams of BPS particles. Notice that this relates the matching of the 8d corrections in dual pictures to the matching of the BPS spectra of the 9d theories. Therefore the matching of our results (and those in \cite{Billo:2009di}) simply follows from heterotic-type I' duality in 9d, which is well understood at the BPS level \cite{Bergman:1997py}. This viewpoint will be useful in section \ref{polyinstanton}.

\subsection{Gravitational couplings}
\label{gravitational}

In this section we compute gravitational and mixed corrections from one-loop diagrams of BPS particles. For concreteness we focus on non-perturbative D0-brane states, although clearly the perturbative terms can be obtained similarly.

Gravitational $R^4$ or mixed $R^2F^2$ corrections can be obtained by replacing the $\tr_{\bf R} F^4$ insertion in the loop diagram by suitable couplings to the external source of field strength or curvature. This can most easily be done by computing in a background source, and subsequently picking the corresponding term in the Taylor expansion in the background. The computation in a background is exactly as the one used in the computation of anomaly polynomials, hence we may borrow the results of contributions from different kinds of fields, which we gather in Appendix \ref{anomaly}. Note that for previously computed pure gauge corrections, the $\tr_{\bf R} F^4$ term can be recovered from the corresponding term in the expansion of the Chern character $\tr_{\bf R} e^{F}$.

Let us compute the non-perturbative $\tr R^4$ correction in the $SO(16)^2$ model. The D0-branes in the boundary are in vector multiplets, and contain states of spin $1/2$ contributing to gravitational couplings via the A-roof polynomial (\ref{aroof}). Taking into account their $SO(16)^2$ multipliticies, they contribute in the following way,
\begin{eqnarray}
\Delta^{R^4}_{s=\frac{1}{2}} = \frac{1}{(4\pi)^4}\frac{1}{360} \, \tr R^4 \, \sum_{n,m}\, \frac{1}{m} \, \left[ \, 240 \, q^{2n\, m}\, +\, 256\,  q^{(2n-1)m }\right] ~.
\end{eqnarray}
There are also contributions from bound states of $2n$ D0-branes in the bulk (as counted in the covering space). They contain massive spin $3/2$ states, which from (\ref{spin32}) contribute as
\begin{eqnarray}
\label{16R43/2}
\Delta^{R^4}_{s=\frac{3}{2}}=\frac{1}{(4\pi)^4}\, \frac{248}{360}\,  \tr R^4 \, \sum_{n,m}\, \frac{1}{m} \, q^{m(2n)}~.
\end{eqnarray}
The total contribution is thus
\begin{eqnarray}
\Delta^{R^4}_{SO(16)}=\frac{1}{(4\pi)^4}\, \frac{1}{360}\,  \tr R^4 \, \sum_{k}\sum_{\ell \large| k}\, \frac{1}{\ell}\, \left[\, 232\, q^{2k}\, +\, 256 \,q^k \,\right]~.
 \end{eqnarray}
This agrees with the heterotic string result in \cite{Gutperle:1999dx}.

Let us also compute the full non-perturbative gravitational and mixed corrections in the $SO(8)^4$ model. Focusing on mixed $\tr R^2 \tr F^2$ corrections for a fixed $SO(8)$ factor, there are contributions only from boundary D0-branes charged under it. Using the relevant term in (\ref{1/2}), the contribution is
\beqa
\Delta_{SO(8)}^{R^2F^2}&=& -\frac 16 \, \frac{1}{(4\pi)^4}\,  \tr R^2 \left(\, 
\tr_{\bf 28} F^{2} \, \sum_{n,m}\, \frac{1}{m}\, q^{2n\, m} \, +\, 8\, \tr_{\bf 8_v} F^2\,  \sum_{n,m}\, \frac{1}{m}\, q^{2n\, m} \, (-1)^m\,  \right. \nn\\
&&\left. \, ~~~~~~~~+8\,\tr_{\bf 8_s} F^2\,  \sum_{n,m}\, \frac{1}{m}\, q^{(2n-1)\, m}\, +\, 8\,\tr_{\bf 8_c} F^2\,  \sum_{n,m}\, \frac{1}{m}\, q^{(2n-1)\, m} (-1)^m\, \right) \nn\\
&=& -\frac{1}{(4\pi)^4}\,  \tr R^2\, \tr F^{2}\, \sum_{k}\sum_{\ell \large| k} \,\frac{1}{\ell}\, q^{2k}~.
\label{mixedso8}
\eeqa
For purely gravitational couplings, there are contributions from all D0-brane states, both from those bound to the D8-branes and from those in the bulk. Taking into account their multiplicites, and their spin components, the contribution is
{\small
\beqa
\Delta_{SO(8)}^{R^4}&=& \frac{1}{(4\pi)^4} \left( \frac{1}{360}\,\tr R^4 +\frac{1}{288}\, (\tr R^2)^2\right)\times \nn \\
&&\times \Bigl(\sum_{n,m}\, \frac{1}{m} \, q^{2n\, m}\, [\, 112\, +\, 128\, (-1)^m ] 
+ \sum_{n,m}\, \frac{1}{m}\, q^{(2n-1)m}\, [\, 128\, +\, 128 (-1)^m] \Bigr)\nn \\
&& + \,\frac{1}{(4\pi)^4} \left( \frac{248}{360}\,\tr R^4 +\frac{56}{288}\, (\tr R^2)^2\right)\, \sum_{n,m}\, \frac{1}{m}\, q^{2n\, m}\, \nn\\
&=&  \,\frac{1}{(4\pi)^4} \left(\,\tr R^4 +\frac{7}{12}\, (\tr R^2)^2\right)\, \sum_k \sum_{\ell |k}\, \frac{1}{\ell}\, q^{2k}\,
\label{gravso8}
\eeqa
}
The results (\ref{mixedso8}) and (\ref{gravso8}) agree with the heterotic and type IIB computations in \cite{Billo:2009di}, up to rescalings of traces and field strengths (which do not modify the agreement for the pure gauge correction).

\section{More general configurations and the prepotential}
\label{moregeneral}

\subsection{General Wilson lines and D8-brane positions}
\label{wilsonpositions}

An interesting generalization corresponds to considering the type IIB orientifold with D7-branes at more general positions in the transverse 2-plane, as encoded in the vevs of complex scalars in vector multiplets. In the type I' picture, a real component of these scalars corresponds to turning on more general D8-brane Wilson lines along the $\IS^1_9$ wrapped by the D0-branes. These degrees of freedom are complexified by considering general positions of the D8-branes, away from the O8-plane. In this section we discuss these generalizations, which turn out to be very simple in our language.

Let us start by considering the configuration with 16 D8-branes on top of each O8-plane, with general Wilson lines along the $\IS^1_9$. We focus on non-perturbative gauge corrections from D0-branes, which are generated only for gauge bosons arising from a single $SO(16)$. The gauge group is generically broken to $U(1)^8$, and we denote by $\phi_i$ and $F_i$ the Wilson line and field strength for the $i^{th}$ $U(1)$ factor. The $SO(16)$ weight vector $\Lambda$ of a D0-brane state encodes its $U(1)^8$ charges, so its contribution follows from (\ref{master}), namely
\beqa
\Delta^{D0}_{F^4}=\frac{1}{4!} \sum_{i,j,k,l}\, \sum_m\,\frac{1}{m}\, q^{n\,m}\, e^{2\pi i \, m\, \Lambda_i\cdot \phi_i}\, \Lambda_i\, \Lambda_j\, \Lambda_k\,\Lambda_l \, F_i F_j F_k F_l
\eeqa
It is straightforward to realize that the above results for $SO(8)$ are a particular case.

It is also easy to include general D8-brane positions away from the $SO(16)^2$ point.  
The basic observation \cite{Bergman:1997gf,Bergman:1997py} is that as a D8-brane is moved away from the O8-plane, the BPS D0-branes stuck on it grow fundamental strings joining them and the dislocated D8-brane, in a way dictated by charge conservation. This string creation effect is dual to the Hanany-Witten brane creation effect \cite{Hanany:1996ie}, and is responsible for an increase in the mass of the 9d BPS state. More explicitly, denoting $\varphi_i$ the $i^{th}$ D8-brane position, the mass of a bound state of $n$ D0-branes with $SO(16)$ weight $\Lambda$ is shifted as $n\to n+\Lambda_i\varphi_i$. Therefore its contribution to the $F^4$ couplings (allowing simultaneously for general Wilson lines) can be expressed as
\beqa
\Delta^{D0}_{F^4}=\frac{1}{4!} \sum_{i,j,k,l}\, \sum_m\,\frac{1}{m}\, e^{2\pi i\, \tau \, m \,n}
\, e^{2\pi i \, m\, \Lambda_i\cdot \Phi_i}\, \Lambda_i\, \Lambda_j\, \Lambda_k\,\Lambda_l \, F_i F_j F_k F_l
\eeqa
where the complex scalars $\Phi=\phi+\tau \varphi$ correspond the complexification of D8-brane positions and $\IS^1_9$ Wilson lines. This complexification will be manifestly geometric in the M-theory perspective in Section \ref{mtheory}. The above expressions can be encoded in a generating functional \cite{Fucito:2009rs}, the 8d prepotential $F(\Phi)$, by promoting the complex scalars to supermultiplets
\beqa
\Phi_i\, =\, \Phi_i\, + \, \theta \gamma^{\mu\nu}\, \theta \, F_{i,\mu\nu}
\label{superfield-gauge}
\eeqa
and writing, modulo constants,
\beqa
\int\, d^8\theta \, F(\Phi)\, =\, \int \, d^8\theta\, \sum_m\,\frac{1}{m^5}\, e^{2\pi i\, \tau n}
\, e^{2\pi i \, m\, \Lambda_i\cdot \Phi_i}\ \, =\, 
\int \, d^8\theta\, \sum_m\,\frac{1}{m^5}\, e^{2\pi i\, \tau n} \, \tr_{\bf R} e^{2\pi i m \Phi}
\quad
\label{schwinger1}
\eeqa
where $\Phi$ is the background for a supermultiplet in the representation ${\bf R}$.
This can be regarded as the computation of a Schwinger one-loop vacuum diagram in the presence of a background for the superfields (\ref{superfield-gauge}). Schematically it is the trace of the operator $e^{2\pi i \Phi}$ over the one-particle BPS spectrum of the spacetime theory
\beqa
F(\Phi)\, =\, \tr_{\cal{H}}\, e^{2\pi i \Phi}~.
\eeqa
The diverse $F^4$ terms are recovered from the $4^{th}$ derivative w.r.t the background fields, as discussed in Section \ref{generalities}.

\subsection{General gravitational terms}
\label{generalgravitational}

The idea can be generalized to include curvature terms, using ideas from Section (\ref{gravitational}). We describe a curvature tensor background as in the computation of anomaly polynomials (see e.g. \cite{AlvarezGaume:1985ex})
\beqa
\frac{R}{2\pi}\,=\,\diag (\epsilon_1,\epsilon_2,\epsilon_3,\epsilon_4) \otimes 
 \begin{pmatrix} & 1  \cr -1 & \end{pmatrix}
\eeqa
with Pontryagin classes given by 
\beqa
\frac{1}{(2\pi)^2} \tr R^2\, =\,\sum_{i<j}\, \epsilon_i^2\epsilon_j^2 \quad ,\quad  
\frac{1}{(2\pi)^4} \tr R^4\, =\sum_{i<j<k<l} \epsilon_i^2\epsilon_j^2\epsilon_k^2\epsilon_l^2. 
\eeqa
Following \cite{Fucito:2009rs} we simply promote the backgrounds $\epsilon_\ell$ to supermultiplets $W_\ell$ similar to (\ref{superfield-gauge})
\beqa
\epsilon_\ell \, \to\, W_\ell \, =\, G_\ell\, +\, R_{\mu\nu}^\ell \, \theta\gamma^{\mu\nu}\theta
\eeqa
with $G_i$ and $R^i_{\mu\nu}$ the graviphoton and curvatures associated to the $i^{th}$ Cartan generator in the Lorentz group.

The prepotential is given by a trace over the one-particle BPS spectrum of the anomaly polynomial operator, regarded as a function of supermultiplets. For instace, for boundary D0-branes, 
\beqa
F(\Phi,W)\, =\, \tr_{\cal{H}}\, e^{2\pi i \Phi} \, \hat{A}(W)
\eeqa 
and all gauge and gravitational corrections are obtained from $\int d^8\theta \, F(\Phi_i,W_\ell)$.

\section{The M-theory point of view}
\label{mtheory}

The non-perturbative contributions are due to D0-branes. These states admit a simple interpretation in the Horava-Witten M-theory lift, as momentum modes of the $E_8$ vector multiplets on the boundaries (as used to derive the type I' D0-brane bound state multiplicities). The non-perturbative contribution should therefore admit a simple description as a one-loop diagram of massless $E_8$ fields in the 10d boundary compactified on $\IT^2$ down to 8d. This description makes the modular properties of the result manifest. It also allows for generalizations of the computation in models with general Wilson lines, not necessarily related to perturbative type I'.

\subsection{The one-loop diagram}
\label{mtheoryoneloop}

Consider Horava-Witten theory on a 2-torus, i.e. M-theory on $\mathbf{R}^8 \times \IS^{1}_{(9)}\times (\IS^{1}_{(10)} / \mathbf{Z}_2) \times \IS^{1}_{(11)}$. We will calculate the 1-loop amplitude of massless $E_8$ gauge bosons with 4 external insertions of gauge field strengths (or curvature tensors, in which case we also include 1-loop diagrams of bulk gravitons).

We begin by studying the massless $E_8$ gauge bosons which live at  10-dimensional boundaries of the $\mathbf{S^{1}_{(10)}} / \mathbf{Z}_2$ interval. Since these particles are stuck at the boundaries we only need to sum over the KK-momenta they carry in the $ \mathbf{ S^{1}_{(9)}\times  S^{1}_{(11)} = T^{2}_{(9,11)}}$ directions,
\begin{eqnarray}
\label{Mgauge}
\mathcal{A}^{gauge}  &=& \frac{1}{4!}
\int_{0}^{\infty} \frac{dt}{t} t^4 \sum_{\ell_I} \int d^{8}\mathbf{p} ~ e^{-\pi t \left( \mathbf{p}^2+G^{IJ} \widetilde{\ell}_{I} \widetilde{\ell}_{J}\right)} \nn\\ 
&=&\frac{1}{4!}
 \int_{0}^{\infty} \frac{dt}{t} \sum_{\ell_9 , \ell_{11}}~ e^{-\pi t  \frac{1}{V_{(2) }\tau_2} | \widetilde{\ell}_{9} -\tau\widetilde{\ell}_{11} |^2}
\end{eqnarray}
where we have denoted, 
\begin{equation}
\label{ }
\widetilde{\ell}_{I} =
\begin{pmatrix}
  \widetilde{\ell}_{9}     \\
  \widetilde{\ell}_{11}   
\end{pmatrix}
=
\begin{pmatrix}
 \ell_9 - \mathbf{\Lambda}\cdot \mathbf{A}_9     \\
 \ell_{11}   - \mathbf{\Lambda} \cdot \mathbf{A}_{11}
 \end{pmatrix}
 \end{equation}
where $ \mathbf{\Lambda}$ denotes the weight vectors of the adjoint ($\mathbf{248}$) of $E_8$ and $\mathbf{A}_{I}$ denotes the Wilson lines along the $I=9,11$ directions of the $\mathbf{T^{2}_{(9,11)}}$. The action of the modular group is manifest in this expression, so the invariance group of the result is the subgroup of $SL(2,\IZ)$ preserving the Wilson line structure. This will be discussed in section \ref{modular}.

Performing a  Poisson resummation on the KK momenta $\ell_9$, we get a sum over  winding numbers $w_9$, 
\begin{eqnarray}
\label{Mgauge2}
\mathcal{A}^{gauge} &=& \frac{\sqrt{\pi}}{4!} 
\int_{0}^{\infty} \frac{dt}{t} t^{-1/2} \sum_{w_9 , \ell_{11}} 
e^{- \frac{\pi^2  w_{9}^{2} }{t} -  \tau_{2}^{2} \widetilde{\ell}_{11}^{2}  t   }  ~
e^{2\pi i w_9 \widetilde{\ell}_{11} \tau_1} e^{2\pi i w_9  \mathbf{\Lambda}\cdot \mathbf{A}_{9}} ~.
\end{eqnarray}

In what follows we consider the Wilson lines $\mathbf{A}_{11}$ to implement the breaking $E_8\to SO(16)$, so as to connect with the type I' description in previous sections. In the above expression, it is convenient to split off the contribution from $w_9=0$, which corresponds to the tree level term from the type I' perspective. After Poisson resummation on $\ell_{11}$ and integration over $t$ it becomes 
\begin{eqnarray}
\label{treegauge}
\mathcal{A}^{gauge}_{(w_{9}=0)}
&=& \frac{\tau_2}{4!\pi} 
 \sum_{ w_{11}\neq 0} \frac{1}{w_{11}^{2}}e^{2\pi i w_{11}\mathbf{\Lambda}\cdot \mathbf{A}_{11} }  
\end{eqnarray}
where we exclude the divergent $w_{11}=0$ term (which is absent in a global tadpole free theory). Splitting ${\mathbf{248}}\to{\bf {120}}+{\bf {128}}$, reintroducing the external $F$ insertions and summing over weights we obtain
\begin{eqnarray}
\label{treegaugeI}
\Delta^{\mathrm{Gauge}}_{w_{9}=0} &=&\frac{2\tau_2}{4!\pi} \left[  \sum_{ w_{11}> 0} \frac{1}{w_{11}^{2}} \, \tr_{\bf 120} F^4\, +  \sum_{ w_{11}> 0} \frac{(-1)^{w_{11}}}{w_{11}^{2}}\,  
\tr_{\bf 128} F^4 \right] \nn \\
&=& \frac{4\pi}{4!} \tau_2\, \tr F^4
\end{eqnarray} 
where in the last line we have used  $\sum_{ w_{11}> 0} \frac{1}{w_{11}^{2}}=\pi^2/6$ and $\sum_{ w_{11}> 0} \frac{(-1)^{w_{11}}}{w_{11}^{2}} =-\pi^2/12$, and the trace identities (\ref{SO16}). Although the result is independent of the Wilson line $\mathbf{A}_9$, it is understood that if $SO(16)$ is broken into several factors, each receives a contribution of this form. This reproduces the tree level $F^4$ coupling for D7-branes, as announced.

The non-zero $w_9$ contribution in (\ref{Mgauge2}) becomes
\begin{eqnarray}
\label{gaugecorr}
\mathcal{A}^{gauge}_{(w_{9}\neq 0)}
=\frac{1}{4!}
\sum_{{}^{w_9 \neq 0}_{\ell_{11} \in {\mathbf Z}}} \frac{1}{|w_9 | } 
e^{-2\pi \tau_2 |w_9 (\ell_{11}-\mathbf{\Lambda}\cdot \mathbf{A}_{11})|} ~
e^{2\pi i \tau_1 w_9 (\ell_{11}-\mathbf{\Lambda}\cdot \mathbf{A}_{11})} ~
e^{2\pi i w_9 \mathbf{\Lambda}\cdot \mathbf{A}_{9}} ~.
\end{eqnarray}
Splitting ${\mathbf{248}}\to{\bf {120}}+{\bf {128}}$, corresponding to $\mathbf{\Lambda\cdot A}_{11}$ being in $\IZ$ or $\IZ+\frac 12$,  and with suitable relabelings of $\ell_{11}$, the contribution can be recast as
\beqa
\label{gaugenp}
\Delta^{\mathrm{Gauge}}_{\rm non-pert} & = & \frac{2}{4!}\, \sum_{{}^{w_9 > 0}_{\ell_{11} > 0}} \frac{1}{w_9 } \, q^{w_9 \ell_{11}}~ \tr_{\mathbf{120}}\,(\, F^4 \, e^{2\pi i w_9 \mathbf{\Lambda}\cdot \mathbf{A}_{9}}\, )\, +\, \mathrm{c.c.}\, \nn\\
&&+ \frac{2}{4!} \sum_{{}^{w_9 > 0}_{\ell_{11} >0}} \frac{1}{w_9 } \,
q^{w_9 (\ell_{11}-1/2)} ~\tr_{\mathbf{128}}\,(\, F^4\,
e^{2\pi i w_9 \mathbf{\Lambda}\cdot \mathbf{A}_{9}}\, ) + \mathrm{c.c.}
\eeqa
where we have reintroduced the $F^4$ insertion. The factor of 2 arises from summing over negative $w_9$. We have also removed the contribution from $\ell_{11}=0$, which corresponds to the massless $SO(16)$ gauge bosons. It can be considered in the perturbative type I' sector, together with the contributions from winding open strings. The latter should arise from wrapped M2-branes in the Horava-Witten theory, whose contribution cannot be reliably computed in M-theory. 

The above amplitude, as already indicated in the subindex, reproduces the non-perturbative D0-brane contributions in type I' in section \ref{computation}. The precise match requires redefining $q\to q^2$ in (\ref{gaugenp}), in order to change the unit of D0-brane charge from the quotient space convention (implicit in the Horava-Witten picture) to the covering space (used in the type I' picture in previuos sections). The above expression reproduces the type I' result for general Wilson lines $\mathbf{A}_9$. 
A slight generalization, allowing for deviations of $\mathbf{A}_{11}$ from the $SO(16)$ case can easily be shown to reproduce the type I' contribution for general D8-brane positions in the interval. Hence the complexification of D8-brane positions and Wilson lines is naturally geometrized in terms of complex Wilson lines in the M-theory setup.

\subsection{The $SO(8)^4$ modular group}
\label{modular}

From the M-theory perspective, the modular group $SL(2,\mathbb{Z})$ of type IIB has a natural geometrical interpretation, and the invariance subgroup of our setups is that preserved by the Wilson lines $\mathbf{A}_9$, and $\mathbf{A}_{11}$. In this subsection we explicitly compute this invariance subgroup for the $SO(8)^4$ model and recover results from the literature. In order to do so, it is convenient to use dimensional regularization to parametrize the infrared divergencies caused by massless states running in the loop. This allows us to derive an alternative expression of the D0-brane 1-loop amplitudes \eqref{Mgauge} in terms of non-holomorphic Eisenstein series, which make the modular properties of the model manifest. Let us compute these gauge amplitudes in $8+2\epsilon$ dimensions and perform a Poisson resummation over both KK-momenta $\ell_I$, instead of just $\ell_9$:
\begin{eqnarray}
\mathcal{A}^{gauge}  &=& \frac{1}{4!}\,V_{(2)}^{\epsilon}
\int_{0}^{\infty} \frac{dt}{t} t^4 \sum_{\ell_I} \int d^{8+2\epsilon}\mathbf{p} ~ e^{-\pi t \left( \mathbf{p}^2+G^{IJ} \widetilde{\ell}_{I} \widetilde{\ell}_{J}\right)} \nn\\
&=&\frac{1}{4!}\,V_{(2)}^{\epsilon} 
 \int_{0}^{\infty} \frac{dt}{t^{1+\epsilon}} \sum_{\ell_9 , \ell_{11}}~ e^{-\pi t  \frac{1}{V_{(2) }\tau_2} | \widetilde{\ell}_{9} -\tau\widetilde{\ell}_{11} |^2} \nn\\
&=&\frac{\Gamma(1+\epsilon)}{4!\,\pi^{1+\epsilon}} 
\sum_{(w_9,w_{11})}\frac{\tau_2^{1+\epsilon}}{|w_9+\tau w_{11}|^{2+2\epsilon}}\,
e^{2\pi i(w_9 \mathbf{\Lambda}\cdot \mathbf{A_9}+w_{11}\mathbf{\Lambda}\cdot \mathbf{A_{11}})}.
\label{Mgauge22}
\end{eqnarray}
where it is here and in the following implied that the term $(w_9,w_{11})=(0,0)$ is excluded in the sum. The prefactor $V_{(2)}^{\epsilon}$ involving the volume of the torus $\mathbf{T^{2}_{(9,11)}}$ has been consistently included for dimensional reasons. For $\mathbf{\Lambda}\cdot \mathbf{A_9}=\mathbf{\Lambda}\cdot \mathbf{A_{11}}=0$ the expression in (\ref{Mgauge22}) is the non-holomorphic Eisenstein series of order $1+\epsilon$, $E_{1+\epsilon}(\tau)$, which has a pole in $\epsilon$. The physical reason for this divergence is in our case that there are massless states running in the loop.


To analyze this expression for the $SO(8)^4$ model we just need to consider the corresponding Wilson lines and their effect in the $\mathbf{248}$ states \eqref{SO(8)reps}. The corresponding contributions are
\begin{eqnarray}
(\mathbf{28,1}): \,\, \Delta^{gauge}_{(\mathbf{28,1})} &=& 
                        \frac{1}{4!\,\pi}\sum_{(w_9,w_{11})} \frac{\tau_2^{1+\epsilon}}{|w_9+\tau w_{11}|^{2+2\epsilon}}\,
                        \mathrm{Tr}_{(\mathbf{28,1})} F_{SO(8)}^4 =
                         \frac{3}{4!\pi}\,E_{1+\epsilon}(\tau)\, (\mathrm{tr}\, F^2)^2 \nn\\
(\mathbf{8_v,8_v}): \,\, \Delta^{gauge}_{(\mathbf{8_v,8_v})} &=&
                        \frac{1}{4!\pi}\sum_{(w_9,w_{11})} \frac{(-1)^{w_9}\tau_2^{1+\epsilon}}{|w_9+\tau w_{11}|^{2+2\epsilon}}\,
                        \mathrm{Tr}_{(\mathbf{8_v,8_v})} F^4 \nn \\
                        &=& \frac{8}{4!\pi}\,[E_{1+\epsilon}(\tau/2)-E_{1+\epsilon}(\tau)]\, \mathrm{tr}\, F^4 \nn\\
(\mathbf{8_s,8_s}): \,\, \Delta^{gauge}_{(\mathbf{8_s,8_s})} &=&
                        \frac{1}{4!\,\pi}\sum_{(w_9,w_{11})} \frac{(-1)^{w_{11}}\tau_2^{1+\epsilon}}{|w_9+\tau w_{11}|^{2+2\epsilon}}\,
                        \mathrm{Tr}_{(\mathbf{8_s,8_s})} F^4 \nn\\
                        &=& \frac{1}{4!\pi}\,[E_{1+\epsilon}(2\,\tau)-E_{1+\epsilon}(\tau)]\,(-4 \mathrm{tr}\, F^4 +3 (\mathrm{tr}\, F^2)^2 - 96 \mathrm{Pf} F) \nn 
\end{eqnarray}
\begin{eqnarray}
(\mathbf{8_c,8_c}): \,\, \Delta^{gauge}_{(\mathbf{8_c,8_c})} &=&
                        \frac{1}{4!\,\pi}\sum_{(w_9,w_{11})} \frac{(-1)^{w_9+w_{11}}\tau_2^{1+\epsilon}}{|w_9+\tau w_{11}|^{2+2\epsilon}}\,
                        \mathrm{Tr}_{(\mathbf{8_c,8_c})} F^4 \nn \\ 
                        &=&\frac{1}{4!\pi} [2 E_{1+\epsilon}(\tau)-E_{1+\epsilon}(2\,\tau)-E_{1+\epsilon}(\tau/2)]\, (-4 \mathrm{tr}\, F^4 +3 (\mathrm{tr}\, F^2)^2 + 96 \mathrm{Pf} F), \nn
\end{eqnarray}
where, for convenience, we have included in the definition of the non-holomorphic Eisenstein series, a factor of $\frac{\Gamma(1+\epsilon)}{\pi^{\epsilon}}$:
\begin{equation}
E_{1+\epsilon}=\frac{\Gamma(1+\epsilon)}{\pi^{\epsilon}} \sum_{(w_9,w_{11})}\frac{\tau_2^{1+\epsilon}}{|w_9+\tau w_{11}|^{2+2\epsilon}}
\end{equation}

In order to relate these expressions to the heterotic results in \cite{Billo:2009di} we need to make the change of variables $\tau\, \to\, -1/(2\tau)$. 
Collecting the terms, we obtain the following total contribution,
\begin{eqnarray}
\Delta^{gauge} &=& \frac{12}{\pi\, 4!}\, [E_{1+\epsilon}(4\tau)- E_{1+\epsilon}(2\,\tau)] \mathrm{tr} F^4 + 
                    \frac{3}{\pi\, 4!} \,[2\,E_{1+\epsilon}(2\,\tau)- E_{1+\epsilon}(4\tau)] (\mathrm{tr} F^2)^2  \nonumber \\
                    &&+ \frac{96}{\pi\, 4!} \,[3\, E_{1+\epsilon}(2\,\tau) -2\,E_{1+\epsilon}(\tau) - E_{1+\epsilon}(4\tau)]\, \mathrm{Pf} F\nonumber\\
                   &=& \frac{12}{\pi\, 4!}\, [E_{1+\epsilon}(4\tau)- E_{1+\epsilon}(2\,\tau)] \mathrm{tr} F^4 +
                    \frac{3}{\pi\, 4!} \,[2\,E_{1+\epsilon}(2\,\tau)- E_{1+\epsilon}(4\tau)] (\mathrm{tr} F^2)^2  \nonumber \\
                    &&+ \frac{96}{\pi\, 4!} \,[E_{1+\epsilon}(\tau+\frac{1}{2}) -E_{1+\epsilon}(\tau)]\, \mathrm{Pf} F.
\label{correctamplitude}
\end{eqnarray}
This expression can be compared to the heterotic results in \cite{Billo:2009di} with the aid of the first Kronecker limit formula, which, with our conventions reads
\begin{equation}
E_{1+\epsilon}(\tau)=\frac{\pi}{\epsilon}+\pi \left[ \gamma_E-\log4\pi-\log(\tau_2|\eta(\tau)|^4)\right]+ O(\epsilon),
\label{Kron}
\end{equation}
where $\gamma_E$ is the Euler-Mascheroni constant. \eqref{correctamplitude} coincides with the heterotic results if we simply drop the divergent term in \eqref{Kron}, i.e. if we substitute every Eisenstein series of order $1+\epsilon$ in \eqref{correctamplitude} by a renormalized Eisenstein series of order 1 defined by\footnote{For further details on this renormalization and its relation to other regularization schemes see \cite{Ghilencea:2002ff}, \cite{Kiritsis:1997em}.}:
\begin{equation}
\hat{E}_1(\tau)\equiv \lim_{\epsilon\to 0} \left[E_{1+\epsilon}(\tau) - \frac{\pi}{\epsilon}-\pi(\gamma_E-\log4\pi) \right]
\end{equation}
This renormalization procedure is manifestly consistent with modular invariance. Note also that in practice the renormalization is only necessary for the $(\mathrm{tr} F^2)^2$ term in \eqref{correctamplitude}, which is the only divergent one, in agreement with the results of section \ref{computation}.

The alternative expression \eqref{correctamplitude} is useful to determine the modular invariance group of the effective action. It is simply made up of $SL(2,\mathbb{Z})$ transformations
\begin{equation}
\tau \to \frac{a\tau +b}{c\tau + d}
\end{equation}
which leave the renormalized Eisenstein functions $\hat{E}_1(\tau)$, $\hat{E}_1(2\tau)$ and $\hat{E}_1(4\tau)$ invariant. Note that the modular properties of renormalized and unrenormalized  Eisenstein functions are the same so we will work in the following with the latter. Since $E_1(\tau)$ is modular invariant, and the invariance group of $E_1(4\tau)$ is a subgroup of the invariance group of $E_1(2\tau)$, we only need to look at the function $E_1(4\tau)$, which transforms as
\begin{eqnarray}
E_1(4\,\tau) \to 
&& \sum_{(w_9,w_{11})} \frac{1}{|c\tau+d|^2}\frac{4\,\tau_2}{|w_9+\frac{a\tau+b}{c\tau+d}\,4w_{11}|^2} \nonumber \\
&& \sum_{(w_9,w_{11})} \frac{4\,\tau_2}{|(dw_9+4bw_{11})+(\frac{c}{4}w_9+aw_{11})\,4\,\tau|^2} .
\end{eqnarray} 
Clearly, this function is only invariant if $c=4n$ ($n\in \mathbb{Z}$), i.e. under transformations of the form
\begin{equation}
\left(
\begin{array}{cc}
a & b \\
4\,n & d
\end{array} \right) \in SL(2,\mathbb{Z})~.
\end{equation}
Hence we have recovered the known result that the effective action is invariant under the subgroup  $\Gamma_0(4) \in SL(2,\mathbb{Z})$.

\section{Polyinstanton effects}
\label{polyinstanton}

We have seen that the $F^4$ and $R^4$ terms of the 8d theory can be obtained as a one-loop computation, in terms of the spectrum of 9d BPS one-particle states in vector (short) multiplets, in an 8d analog of \cite{Harvey:1995fq}. In the type I' model, such BPS states are fundamental strings or D0-branes, leading to perturbative or non-perturbative corrections, with the latter reproducing elegantly the D$(-1)$-brane instanton sums of the T-dual type IIB orientifold. In the heterotic dual, the 9d BPS particles are winding and momentum states of fundamental strings, and the one-loop diagram reproduces the worldsheet instanton contributions computed in the literature. The one-loop description therefore shows that the agreement of the 8d corrections in type II-heterotic duals follows from the agreement of the 9d spectrum of one-particle BPS states in heterotic-type I' duality, which has been extensively studied in \cite{Bergman:1997py}. 

\subsection{Polyinstantons in 8d}
\label{8dpoly}

It is worthwhile to note that the contribution from a single 9d BPS D0-brane can correspond to a multi-instanton contribution on the type IIB side. This is particularly manifest for BPS D0-brane bound states of $k$ elementary D0-branes. This is the 8d analog of a similar phenomenon in the 4d $\cn=2,1$ context. The fact that multiple instantons can conspire to contribute to the non-perturbative $\cn=1$ superpotential (or $\cn=2$ hypermultiplet metric) \cite{GarciaEtxebarria:2007zv,GarciaEtxebarria:2008pi} was interpreted in \cite{Collinucci:2009nv} as the fact that in the T-dual theory the corresponding BPS particles form a bound state at threshold.

In the 4d setup \cite{Blumenhagen:2008ji} considered a different kind of multiple instanton effect, dubbed polyinstanton, which also has an 8d analog in our setup.
The polyinstantons in \cite{Blumenhagen:2008ji} were claimed to violate heterotic-type I duality. 
In this section we address this puzzle for 8d polyinstantons, shedding light from a new perspective, valid also in the 4d setup. The bottom line is that polyinstanton processes can be interpreted as reducible Feynman diagrams which do not contribute to the microscopic 1PI effective action.

\begin{figure}[!htp]
\centering
\includegraphics[scale=0.50]{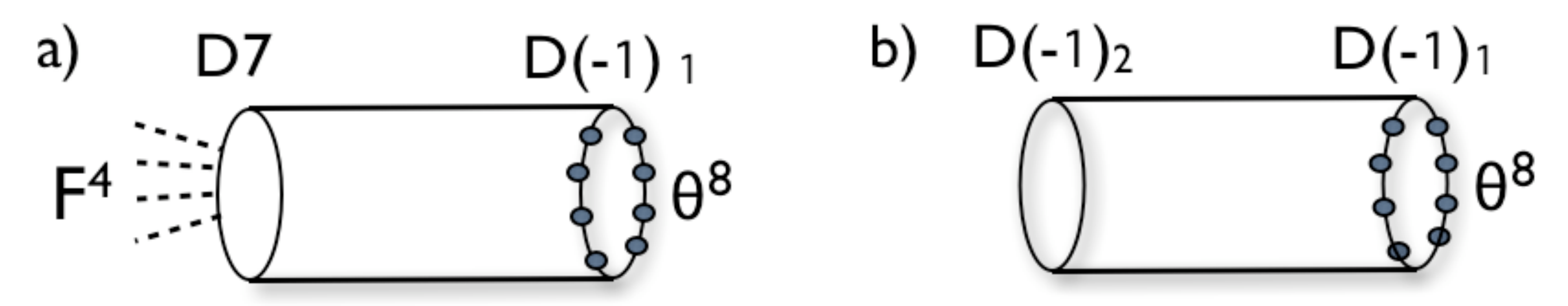}
\caption{\small a) D$(-1)$-brane instanton correction to the $F^4$ coupling on a D7-brane. b) A related diagram describes a D$(-1)$-brane instanton correction to the action of a second D$(-1)$-brane instanton.}
\label{poly}
\end{figure}

Let us start by introducing the 8d polyinstanton corrections to e.g. $F^4$, in complete analogy with the $F^2$ corrections \cite{Blumenhagen:2008ji}. The microscopic diagram leading to a D7-brane $F^4$ correction from a D$(-1)$-brane instanton  includes a cylinder diagram with a boundary on the D7-brane (with 4 fields strength insertions) and a boundary on the D$(-1)$ (with 8 fermion zero mode insertions saturating the instanton Goldstinos), see figure \ref{poly}a. As shown in figure \ref{poly}b, there is a similar diagram, with the D7-brane replaced by a second D$(-1)$-brane instanton, and with no insertions on the correspoding boundary. Labeling the two instantons $1,2$ to avoid confusion,
this diagram represents the correction from D$(-1)_1$ to the action of D$(-1)_2$. Considering now an $F^4$ term induced by D$(-1)_2$, the inclusion of this correction would naively lead to a contribution to the 8d effective action schematically of the form
\beqa
 \int d^8x\, \tr F^4\, e^{-(S_2+e^{-S_1})}\, =\, \int d^8x\, \tr F^4\,  \sum_{n=0}^\infty \, \frac{1}{n!}\, e^{-S_2}\, (e^{-S_1})^n
\label{naive}
\eeqa
Microscopically, the $n^{th}$ term corresponds to a polyinstanton process with one D$(-1)_2$ instanton and $n$ independent D$(-1)_1$ instantons. The zero modes of the latter are saturated through D$(-1)_1$-D$(-1)_2$ cyclinders as in Figure \ref{poly}b.
The result involves an integration over the relative positions of the instantons in the 8d space, just like in the 4d case \cite{Blumenhagen:2008ji}. The contribution is therefore in principle not localized on coincident instantons, as opposed to the multiinstantons studied in \cite{GarciaEtxebarria:2007zv}. In particular, since the polyinstantons in general sit at different locations in the internal space, the saturation of fermion zero modes can take place independently of the distances among instantons in 8d.

\subsection{Polyinstantons and heterotic-type II orientifold duality}
\label{polyduality}

It is straightfoward to use the $F^4$ results in previous sections to compute these effects exactly (i.e. by summing over multiple instantons of each kind), in particular for D$(-1)$-brane instantons sitting on top of D7-branes. However, a general analysis, together 
with the type I' interpretation in terms of the 9d one-particle BPS spectrum, suffices to make the clash with the heterotic result manifest, and to suggest its resolution.

The 8d corrections arising from standard D$(-1)$-brane instantons correspond under T-duality to one-loop diagrams of 9d BPS D0-brane one-particle states. These are directly translated to one-loop diagrams of 9d BPS states in the heterotic dual, reproducing the genus one worldsheet instanton contributions. This contribution in principle includes certain D-brane multi-instantons, namely those T-dual to 9d particles which form BPS bound states at threshold, and whose hallmark is that their contribution is localized on configurations of coincident instantons. 

Polyinstanton processes however involve instantons whose T-dual particles do not combine into 9d one-particle BPS bound states. This is manifest as in general the individual instantons sit at different points in the internal space, and this separation can persist in the type I' dual, e.g. when they map to D0-branes on different $SO(16)$ boundaries. Therefore they are manifestly not included in the one-loop diagram of one-particle BPS states, and hence in the heterotic genus one worldsheet contribution.

There is a clear way out of this potential clash with duality. The heterotic genus one worldsheet diagram (and so the type I' one-loop diagram) computes the one-loop correction to the 1PI action. Namely it includes the effects of massless states (and is hence non-holomorphic) but does not include {\rm reducible} contributions. These can be later generated by computing tree level diagrams using the effective vertices of the 1PI action. We will now argue that D-brane polyinstanton effects actually correspond to such reducible diagrams, and hence do not contribute to the 1PI action, restoring agreement between type II orientifolds and their heterotic duals.

\subsection{Polyinstantons in spacetime as reducible diagrams}
\label{poly1pi}

\begin{figure}[!htp]
\centering
\includegraphics[scale=0.60]{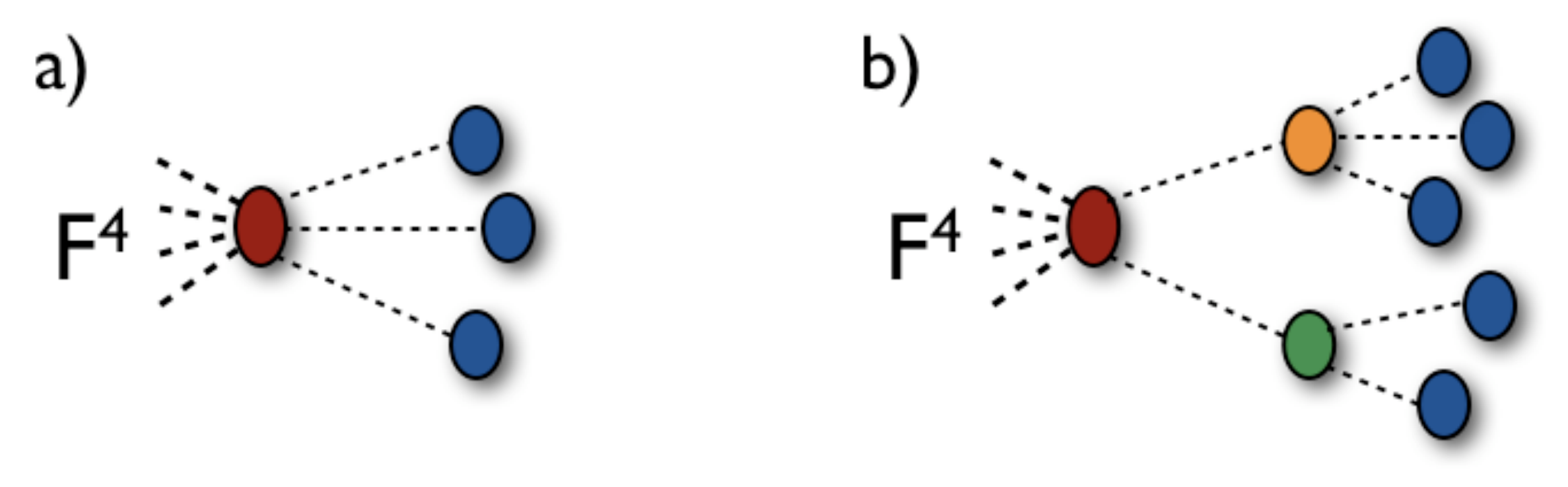}
\caption{\small Polyinstanton processes as reducible spacetime Feynman diagrams. Colored blobs denote elementary instanton interactions in the 1PI action, joined by propagating closed string modes. Summing over polyinstanton processes like a), with arbitrary numbers of blue blobs, reproduces an effective exponential correction to the action of the red blob instanton. Figure b) shows richer polyinstanton processes, involving the elementary interactions described by (\ref{interaction1}) and (\ref{interaction2}).}
\label{reducible}
\end{figure}

In type IIB the polyinstantons correspond to individual D$(-1)$-instantons joined by cylinder diagrams. As the instantons are in general located at (possibly widely) different locations in 8d space, it is natural to interpret the cylinders as a tree level closed string exchange, and the corresponding processes as reducible. Thus polyinstantons do not induce new terms in the microscopic 1PI action, but are rather generated by Wick contractions of other elementary effective vertices in the 1PI action. The picture is particularly clear in the type I' model, where the polyinstanton is given by a Feynman diagram with a loop of BPS particles with four field strength insertions, joined by  closed string propagators to other loops of BPS particles (which can be subsequently joined to other propagators and loops), see Figure \ref{reducible}. 

The exponential combinatorics of polyinstantons in (\ref{naive})  is simply the combinatorics of spacetime Feynman diagrams with two basic kinds of interaction diagrams, see Figure \ref{reducible}a. For massless closed string states, these can be explicitly obtained in the factorization limit. One basic interaction vertex corresponds to an instanton coupling to $F^4$ with emission of $n$ massless closed string states. It simply follows from expanding the $F^4$ instanton corrections in the fluctuations of the  dynamical modulus controling the instanton action, in our case $\tau$. Expanding it into a vev plus fluctuation, $\tau_0+\tau$,  each $F^4$ instanton correction produces terms of the following form,
\beqa
q^N \, \tr F^4\; \to \; \sum_{n=0}^\infty \frac{(2\pi i N)^n}{n!}\, q_0^N \,\tr F^4\, \tau^n~.
\label{interaction1}
\eeqa
The second kind of vertex is the emission of a massless closed string from an instanton. 
Although unfamiliar, this contribution indeed exists, as follows. The fluctuation $\tau$ is a complex scalar belonging to a multiplet whose on-shell structure has the form
\beqa
\Phi_\tau\, =\, \tau \, \, +\, \ldots\, + \theta^8\, \partial^4 {\ov \tau}~.
\eeqa
This follows from the orientifold truncation of the chiral on-shell superfield $\Phi$, satisfying ${\bar D}\Phi=0$, $D^4\Phi={\bar D}^4{\bar\Phi}=0$, in \cite{Green:1998by}. This gives the supersymmetric completion of the instanton action, and the last term
corresponds to an interaction saturating all the instanton goldstino zero modes with one insertion of ${\ov\tau}$. Therefore one generates the couplings
\beqa
\int \, d^8x\,d^8\theta\, e^{2\pi iN(\tau_0+\Phi_\tau)}\, \to\, 
\int \, d^8x\,\, q_0^N\, \, \partial^4{\ov\tau}\, +\, \cdots~.
\label{interaction2}
\eeqa
The term we required has been separated out explicitly. The other terms can be used to construct more involved diagrams, as in Figure \ref{reducible}b.

As we will argue in the next Section, the interpretation of polyinstantons as reducible diagrams remains valid in 4d with minor modifications, due only to the lower supersymmetry which leads to a reduced number of instanton fermion zero modes. The relevant cylinder diagram stretching between the different instantons saturates 4 fermion zero modes on a boundary, while the other remains empty. The cylinder can be regarded as a tree exchange of closed strings, and in the factorization limit the two basic interactions are analogous to (\ref{interaction1}), (\ref{interaction2}) with simple modifications: reduction $F^4\to F^2$ and $\theta^8 \partial^4\bar\tau\to \theta^4\partial^2\bar\tau$. Hence the ideas apply to the 4d $\cn=2$ K3 compactifications in Section \ref{compactification}, and also to the 4d $\cn=1$ orbifolds thereof considered in \cite{Blumenhagen:2008ji}, see also \cite{Camara:2008zk}.

The fact that polyinstantons contribute to Wilsonian actions is consistent with many of the physical arguments in \cite{Blumenhagen:2008ji}. For instance, consider a gauge sector on a stack of D-branes whose gauge kinetic function receives non-perturbative corrections from other instantons. If the gauge sector develops a gaugino condensate, its  scale is determined by the full gauge kinetic function (including its exponential corrections).

\subsection{Effective 1PI and Wilsonian actions}

In addition to the instanton generated couplings with a massless closed string field in (\ref{interaction1}) and (\ref{interaction2}), the 1PI action also contains couplings to massive closed string fields. This action, which we use in order to compare one-loop corrections between orientifolds and heterotic, does not integrate out reducible diagrams with massive particle exchange and it should therefore be considered as a microscopic 1PI action. Note however that it is in many cases customary to instead define an {\it effective} 1PI action where, in addition to all irreducible diagrams, also reducible diagrams with exchange of massive particles (but not reducible diagrams with exchange of light particles) are integrated out. Thus, it is the effective 1PI action that is directly related to the Wilsonian one and it is in these two types of actions where the local 
polyinstanton terms, corresponding to reducible diagrams with massive particle exchange, appear. In contrast, because of the fact that these diagrams are reducible, they are not included in the microscopic 1PI action, which is the one we map to the heterotic side.

It is worthwhile to mention that our BPS particle viewpoint provides an
efficient resummation tool for these effects. For instance, consider the instanton generated contribution to $\tr F^4$ terms, with the general structure
\beqa
\tr F^4 \left( \sum_k N_k q(\tau)^k \right)~
\label{single-instanton}
\eeqa
where $N_k = \sum_{N | k} \frac{1}{N}$. This implies a modification of the tree level term $\tau$, or equivalently, of the instanton action to
\beqa
\tau' = \tau + \sum_k N_k q(\tau)^k~.
\eeqa
Thus, the $\tr F^4$ correction including polyinstantons (those with only one level of branches form the 'main' instanton) is obtained by replacing $\tau\to \tau'$ in (\ref{single-instanton}) such that the coefficient of the $\tr F^4$ becomes
\beqa
\label{double}
\sum_k N_k \,q(\tau')^k  &=& \sum_k N_k \,q(\tau)^k\,\exp \left[ 2\pi i k \left(\sum_r N_ r \,
q(\tau)^r\right)\right] \nn \\
&=& \sum_k N_k \, q(\tau)^k +2\pi i  \sum_{k,r} k \, N_k  N_r \, q(\tau)^k q(\tau)^r + \cdots~.
\eeqa
Further iterations would produce contributions from polyinstantons with
more levels of branches from the main instanton. It is however clear already from the structure of the term with a double sum in (\ref{double}) that it is generated by more than one instanton. As discussed above, contributions like (\ref{double}) are not included in the microscopic 1PI action since they would involve reducible diagrams. In other words, when the instantons $k$ and $r$ in (\ref{double}) are T-dualized into D0-particles they do not form a one-particle BPS bound state, but rather a multiparticle state. Instead, the contribution to the microscopic 1PI action of a bound state of $k$ and $r$ D0-particles is already accounted for by (\ref{single-instanton}), with $k$ there playing the role of $k+r$. 

In this picture the polyinstantons on the orientifold side arise from higher loop diagrams. These corrections (or their 4d $\CN=2$ version) do not violate non-renormalization theorems, since these are loop diagrams in the 9d theory. Given the  one-to-one  map between heterotic/orientifold 9d BPS states, we expect a similar higher-loop interpretation of the above terms in the heterotic picture.

\section{Compactification}
\label{compactification}

In this section we consider compactification of the 8d theory on K3,  leading to models with 4d $\CN=2$ supersymmetry. From the type II perspective, the models can be regarded as an orientifold of a K3$\times \IT^2$ compactification. Cancellation of RR tadpoles requires the gauge D-branes to carry a non-trivial gauge bundle with instanton number 24 (assuming no spacetime filling D3-branes in the IIB model, or D4-branes in the type I'). For concreteness, we focus on models where the internal bundle can be embedded in $SO(16)\times SO(16)$, so that in the type I' picture the D8-branes on each boundary carry instanton numbers $(12+n,12-n)$. Focusing on a given $SO(16)$, the K3 gauge bundle with structure group $K$ breaks the gauge group to the commutant $H$. This can be further broken to its Cartan subalgebra by turning on generic D-brane Wilson lines or positions on $\IT^2$. These compactifications are very similar (and often dual) to $E_8\times E_8$ models on K3$\times \IT^2$ 
 studied in e.g. \cite{Aldazabal:1995yw,
Kawai:1995hy,Seiberg:1996vs,Aldazabal:1996du,Henningson:1996jz,LopesCardoso:1996nc,Stieberger:1998yi,Weiss:2007tk}.
The 4d theory contains hypermultiplets describing the K3 moduli and the compactification bundle moduli, and vector multiplets describing the gauge D-brane Wilson lines and/or positions on $\IT^2$. There is a further vector multiplet containing the dilaton, arising from the $\IT^2$ compactification of a 6d tensor multiplet. In certain orientifolds of K3 orbifolds there are additional vector multiplets arising from 6d tensor multiplets, determined by the orientifold projection on twisted sectors \cite{Polchinski:1996ry}. Although such models do not admit a perturbative heterotic dual, the non-perturbative corrections to their effective action can be studied from the type II orientifold side with our present techniques.

Due to the reduction of the supersymmetry, the non-perturbative corrections from BPS instantons with minimal number of fermion zero modes correspond to terms $\tr F^2$, namely contributions to the vector multiplet prepotential, or to gravitational corrections $\tr R^2$, which can be included in a generalized prepotential. D-brane instantons with additional neutral fermion zero modes may contribute to higher F-terms, but we rather focus on the minimal case. Due to the familiar $\CN=2$ decoupling theorems, the corrections are independent of the hypermultiplets, so they are insensistive to the K3 geometry or the bundle moduli. The resulting effective action is therefore fixed in terms of the topological properties of the model, namely the instanton number $n$.

In the type IIB picture, there are perturbative contributions to the prepotential, as well as non-perturbative contributions arising from D$(-1)$-brane instantons and euclidean D3-branes wrapped on K3. In the type I' picture these contributions can be computed as a one-loop diagram of 5d BPS particles. These correspond to either perturbative 9d states (open strings winding along the interval) in quantum groundstates on K3, 9d D0-branes (in the bulk or at the boundaries, and labeled by the D0-brane charge) in quantum groundstates on K3, or genuinely 5d particles arising from D4-branes wrapped on K3. Although all these particles are on equal footing at the level of the 5d BPS spectrum, the BPS multiplicities of the D4-brane particle states and their quantum numbers under the unbroken gauge groups are not known, and we skip their discussion. In what follows we focus on the D0-brane particles, with the discussion of the perturbative states being similar. We also consider trivial 
 Wilson lines, whose further inclusion is straightforward.

The D$(-1)$-brane instanton contribution is T-dual to a one-loop diagram of 5d particles corresponding to the quantum groundstates of D0-branes on K3, with zero momentum on the interval. The 5d BPS degeneracies are determined by the cohomology of the D0-brane quantum mechanics problem on K3. Focusing on corrections to the gauge kinetic function of the 4d gauge group $H$ arising from a 9d $SO(16)$ factor, there are non-perturbative contributions from the D0-branes at the corresponding boundary. These transform in the representations ${\bf 120}$ or ${\bf 128}$, which we denote generically $R_{16}$, so their dynamics is coupled to the K3 gauge bundle. Using the decomposition
\beqa
SO(16) & \to & K\times H \nonumber \\
R_{16} & \to & \sum_i\, (\, R_{K,i},R_{H,i})
\eeqa
a 9d D0-brane state in the $R_{SO(16)}$ produces a number $n_{R_H}$ 5d D0-brane BPS states in the representation $R_H$, given by the index of the relevant Dirac operator
\beqa
n_{R_H} & = & \sum_i\, \int_{\rm K3}\, {\rm Ch}_{R_{K,i}}(F)\, \hat{\rm A}(R)\, \nonumber \\
 & = & \sum_i\, \int_{\rm K3}\, \left(\, {\rm tr}_{R_{K,i}} F^2 +\, {\rm dim}\, R_{K,i}\, \tr R^2\,\right)
\eeqa
The contribution from these 5d particles to the gauge kinetic function of $H$ can be obtained from the general expression (\ref{generalamp}) for $d=5$, $k=2$. Since $d/2-k=1/2$, it can be recast in the form (\ref{1loop1/2}), formally identical (and for good reasons as we will see) to $F^4$ corrections from 9d particles. It is thus given by
\beqa
\Delta_{4d}\, =\, \sum_m \, \frac{1}{m}\, \sum_i\, q^{n_{R_{16}}\, m}\, n_{R_H}\, 
\, \tr_{R_H} F^2 \, 
\label{5dd0s}
\eeqa
Here $n_{R_{16}}=2n,2n+1$ is the D0-brane charge of a 9d state in the $R_{16}={\bf 120}, {\bf 128}$ of $SO(16)$, respectively. The inclusion of general Wilson lines and brane positions is straightforward, following section \ref{wilsonpositions}.
Note that an expression similar to (\ref{5dd0s}) is valid for the contribution of other 5d BPS states, like perturbative states or D4-brane particles, by simply replacing the corresponding instanton exponential weight $q$, and using the appropriate BPS multiplicities $n_{R_H}$. The latter are straightforward to obtain for perturbative states, but are not known for D4-brane particles.

For 5d D0-branes, the expression (\ref{5dd0s}) can be recast as
\beqa
\Delta^{D0}_{4d} & = & \, \sum_m \, \frac{1}{m}\, \sum_i\, q^{n_{R_{16}}\, m}\,
\sum_i\, \int_{\rm K3}\, \left(\, {\rm tr}_{R_{K,i}} F^2 +\, {\rm dim}\, R_{K,i}\, \tr R^2\,\right)\, \int_{4d} \tr_{R_H} F^2 \, \nonumber \\
& = & \sum_m \, \frac{1}{m}\, q^{n_{R_{16}}\, m}\, \int_{{\rm K3} \times M_4}\left( \, \tr_{R_{16}} F^4 \, +\, \tr_{R_{16}} F^2\, \tr R^2\, \right)\,  \nn\\
& = & \sum_m \, \frac{1}{m}\, q^{n_{R_{16}}\, m}\, \int_{{\rm K3}\times M_4}\, {\rm Ch}_{R_{16}} (F)\, {\hat{\rm A}}(R) 
\label{last4d}
\eeqa 
where in the last equality we have included the purely gravitational terms, which are computed similarly, including the contribution from 9d bulk D0-brane states. Eq. (\ref{last4d}) shows that the contribution from 5d D0-brane states can be obtained by simple dimensional reduction on K3 of the $F^4$ corrections in the 8d theory. This follows from the fact that the contributing 5d particles are just groundstates of the 9d particles, so no information is lost in the dimensional reduction truncation. 
From the type IIB viewpoint, the D$(-1)$ instantons of the 4d theory are essentially those of the 8d theory, with 8 fermion zero modes lifted by the interaction with the curvature and gauge bundle on K3. Careful saturation of the latter for the diverse amplitudes should reproduce the prefactor corresponding to the index of the Dirac operator discussed above, which in this language should be regarded as the Euler characteristic of the relevant instanton moduli space. It is interesting that the type I' picture provides an alternative, and very transparent, interpretation of this factor.

The above argument can be repeated for the perturbative type I' contribution. Hence the 8d prepotential directly produces a large part of the 4d $\CN=2$ prepotential (as applied in certain local orbifolds in \cite{Fucito:2009rs}) and many of the properties of the latter are inherited to the former. This applies in particular to our interpretation of the polyinstanton processes in Section \ref{polyinstanton}. There is furthermore no obstruction to applying further freely acting orbifold quotients which reduce the supersymmetry to 4d $\CN=1$, maintaining the basic properties of the instantons and polyinstantons, as in \cite{Blumenhagen:2008ji}.

\section{Conclusions}
\label{conclusions}

In this paper we have studied corrections to quartic gauge and curvature couplings in 8d type I' models, recovering and generalizing results from the heterotic and type IIB sides. The type I' perspective allows to compute these effects as simple one-loop diagrams, with the result determined by the  multiplicities and quantum numbers of the BPS states in the 9d theory.
This provides an interesting generalization of the analysis in \cite{Collinucci:2009nv}, to models containing orientifold planes. It would be interesting to device efficient tools to compute the relevant BPS multiplicities in 4d orientifold compactifications with $\CN=1$ supersymmetry.  This would be an important step towards the systematic computation of non-perturbative superpotentials in 4d theories with four supercharges.

We have shown that the spectrum of 9d bound states codifies important information concerning the nature of multiple D-brane instanton effects on the type IIB side. Namely, loops of 9d bound states are mapped to multi-instanton effects to the 1PI effective action, in which several instantons conspire to cancel their additional zero modes and contribute to BPS protected quantities,  similar to the effects in \cite{GarciaEtxebarria:2007zv,GarciaEtxebarria:2008pi}. On the other hand, processes involving multi-particle states in 9d correspond to polyinstanton effects.
 It is satisfactory that the type I' picture encodes the two subtly different situations in a simple way.

Our analysis is essentially global, in the sense that we include the effects of the full BPS spectrum of the theory. It would be interesting to find particular limits of our computations where one recovers the results for local type IIB models, in order to connect with the useful tools of localization for D-brane instantons, exploited in \cite{Fucito:2009rs}.

We have connected the modular properties of the quartic corrections to a geometric modular group of $\IT^2$ compactification of Horava-Witten theory. An interesting direction would be to use the latter picture to recover quartic corrections in non-perturbative type IIB vacua  \cite{Lerche:1998nx,Lerche:1998gz,Lerche:1998pf},  such as those described by F-theory, which can include exceptional groups. Such corrections may play an interesting role (e.g. for gauge coupling unification) in the recent phenomenological models in F-theory.

We hope to come back to these and other questions in future work.

\newpage

\begin{center}
{\bf Acknowledgements}
\end{center}

It is a pleasure to thank P. G. C\'amara, L.E. Ib\'a\~nez and D. Persson for useful discussions. A.M.U. thanks M. Gonz\'alez for kind encouragement and support, and the CERN TH group for hospitality during completion of this work, and more in general for being home for the last five years. The research of C.P. was supported by an EU Marie Curie EST fellowship. Work by P.S. and A.M.U. has been supported by Plan Nacional de Altas Energ\'{\i}as, FPA-2006-05485, Comunidad de Madrid HEPHACOS S2009/ESP-1473 and by the Ministerio de Ciencia e Innovacion through the grant FPA2009-07908. P.S. acknowledges financial support from Spanish National
Research Council (CSIC) through grant JAE-Pre-0800401, and thanks the Galileo Galilei Institute for Theoretical Physics for the hospitality and the INFN for partial support during the completion of this work.

\newpage

\appendix

\section{Trace structures}
\label{traces}

This appendix provides some trace identities used in the main text. Most can be extracted from \cite{Erler:1993zy}, except the pfaffian contribution for $SO(8)$, which we have directly computed.
\begin{eqnarray}
\label{SO16}
&& {\mathbf{SO(16)}} \nonumber \\
&& \quad \tr_{\mathbf{ 120}} F^{4}_{SO(16)}= 8 \tr F^{4}_{SO(16)}  +3  ( \tr F^{2}_{SO(16)})^2 \nn  \\ 
&& \quad \tr_{\mathbf{ 128}} F^{4}_{SO(16)}= - 8 \tr F^{4}_{SO(16)}  +6  ( \tr F^{2}_{SO(16)})^2 
\end{eqnarray}

\begin{eqnarray}
\label{traceso8}
&& {\mathbf{SO(8)}} \nonumber \\
&& \quad \tr_{\mathbf{8_s}} F_{SO(8)}^2\, =\,  \tr_{\mathbf{8_c}} F_{SO(8)}^2\, =\, \tr F^2 \nn \\
&& \quad \tr_{\mathbf{28}} F^{4}_{SO(8)}\, = \, 3\, (\tr F_{SO(8)}^2)^2 \nn \\
&& \quad \tr_{\mathbf{8_s}} F^{4}_{SO(8)}\, = \, -\frac 12\, \tr F_{SO(8)}^4\, +\,\frac 38\, (\tr F_{SO(8)}^2)^2\, -\, 12 \Pf F \nn \\
&& \quad \tr_{\mathbf{8_c}} F^{4}_{SO(8)}\, = \, -\frac 12\, \tr F_{SO(8)}^4\, +\,\frac 38\, (\tr F_{SO(8)}^2)^2\, +\, 12 \Pf F 
\end{eqnarray}
where 
\beqa
\Pf F= \frac{1}{2^8}\, t_8^{\mu_1\ldots\mu_8}\, \varepsilon_{a_1\ldots a_8} \, F^{a_1a_2}_{\mu_1\mu_2} \cdots F^{a_7a_8}_{\mu_7\mu_8}
\eeqa
where $t_8$ is the Lorentz $SO(8)$ antisymmetric invariant tensor.

\section{Characteristic classes}
\label{anomaly}

The computation of an amplitude with external gauge field strength or curvature insertions can be performed by computing in the presence of a general gauge or curvature background and selecting the appropriate term in a power expansion. The computation  in general backgrounds is a standard tool in the computation of anomalies,  from which we can borrow the results, see e.g. \cite{AlvarezGaume:1985ex}. In those conventions, the Chern character is given by
\begin{equation}
\label{ }
\mathrm{ch}(F)= \tr_{\mathbf{R}}\left[ \exp  \left( \frac{iF}{2\pi}\right)\right]=r + \frac{i}{2\pi}\tr_{\mathbf{R}} F-
\frac{2}{(4\pi )^2}\tr_{\mathbf{R}} F^2- \frac{i}{6(2\pi )^3}\tr_{\mathbf{R}} F^3+\frac{2}{3(4\pi )^4}\tr_{\mathbf{R}} F^4 +\cdots
\end{equation} 
where $r=\tr_{\mathbf{R}}\mathbf{1}$ denotes the dimension of the representation $R$ of the gauge group. The A-roof genus (relevant for spin 1/2 particles) is
\begin{equation}
\widehat{A}(R)=1+ \frac{1}{12(4\pi )^2}\tr R^2 + \frac{1}{(4\pi)^4}\left[ \frac{1}{288}\left( \tr R^2\right)^2 +\frac{1}{360}\tr R^4 \right] +\cdots
\label{aroof}
\end{equation}
while the polynomial relevant for spin 3/2 particles is given by
\begin{equation}
\label{ }
\tr \left[ \exp  \left( \frac{R}{2\pi}\right)\right]=k + \frac{1}{2\pi}\tr R+
\frac{2}{(4\pi )^2}\tr R^2+ \frac{1}{6(2\pi )^3}\tr R^3+\frac{2}{3(4\pi )^4}\tr R^4 +\cdots
\end{equation}
where $k=\tr \mathbf{1}$. Since our D0-particles are in the  $\mathbf{8_v}+\mathbf{8_s}$ of the $SO(8)$ Lorentz group, $k=8$.

In the Horava-Witten picture we get spin 1/2 states from the boundaries and spin 3/2 states from the bulk. The boundary states carry one spinor index and one gauge index (in representation $\mathbf{R}$) and thus the contribution from the spin 1/2 states  is obtained from 
\begin{eqnarray}
\label{1/2}
\widehat{A}(R)~ \mathrm{ch}(F)& = & r - \frac{2}{(4\pi )^2}\tr_{\mathbf{R}} F^2+\frac{r}{12(4\pi )^2}\tr R^2 \nn  \\
 &  &+ \frac{1}{(4\pi)^4}\left[ \frac{2}{3}\tr_{\mathbf{R}} F^4- \frac{1}{6}\tr R^2\tr_{\mathbf{R}} F^2+\frac{r}{288}\left( \tr R^2\right)^2  +\frac{r}{360} \tr R^4 \right] +\cdots \quad\quad \quad
\end{eqnarray}
The contribution from the gravitino and its spin 1/2 partner is obtained from 
\begin{eqnarray}
\label{3/2}
\widehat{A}(R)~\tr \left[ \exp  \left( \frac{R}{2\pi}\right)\right] & = & 8+\frac{17}{6(4\pi )^2}\tr R^2 +  \frac{1}{(4\pi)^4}\left[ \frac{56}{288}\left( \tr R^2\right)^2  +\frac{248}{360} \tr R^4 \right] +\cdots \quad\quad\quad
\label{spin32}
\end{eqnarray}
Note that the spin 3/2 states (bulk) can only give a contribution corresponding to the $2n$-tower since the $(2n-1)$-tower will always have one (unpaired) D0-particle stuck at the  O8-plane.

\end{document}